\documentclass[a4paper,fleqn,usenatbib,useAMS]{mnras}

\usepackage{graphicx}	
\usepackage{amsmath}	
\usepackage{amssymb}	
\usepackage{multicol}        
\usepackage{bm}		
\usepackage{pdflscape}	

\usepackage[T1]{fontenc}
\usepackage{ae,aecompl}

\usepackage{newtxtext,newtxmath}

\title[Genetic Routine for Astro. Period Estimation]{GRAPE: Genetic Routine for Astronomical Period Estimation}

\author[P. R. McWhirter, et al.]{Paul R. McWhirter,$^{1,2}$\thanks{Contact e-mail: \href{mailto:P.R.McWhirter@2014.ljmu.ac.uk}{P.R.McWhirter@2014.ljmu.ac.uk}}{ Iain A. Steele,$^{1}$}{ Abir Hussain,$^{2}$}{ Dhiya Al-Jumeily$^{2}$}\newauthor{and Marley M. B. R. Vellasco$^{3}$}
\\
$^{1}$Astrophysics Research Institute, LJMU, IC2, Liverpool Science Park, 146 Brownlow Hill, Liverpool, L3 5RF, UK
\\
$^{2}$Applied Computing Research Group, Department of Computer Science, LJMU, James Parsons Building, 3 Byrom Street, Liverpool, L3 3AF, UK
\\
$^{3}$Pontif\'icia Universidade Cat\'olica do Rio de Janeiro, R. Marqu\^es de S\~ao Vicente, 225 - G\'avea, Rio de Janeiro - RJ, 22451-900, Brazil}

\date{Last updated XXXX; in original form YYYY}

\pubyear{2018}

\begin{document}
\label{firstpage}
\pagerange{\pageref{firstpage}--\pageref{lastpage}}
\maketitle

\begin{abstract}
Period estimation is an important task in the classification of many variable astrophysical objects. Here we present GRAPE: Genetic Routine for Astronomical Period Estimation, a genetic algorithm optimised for the processing of survey data with spurious and aliased artefacts. It uses a Bayesian Generalised Lomb-Scargle (BGLS) fitness function designed for use with the Skycam survey conducted at the Liverpool Telescope. We construct a set of simulated light curves using both regular and Skycam survey cadence with four types of signal: sinusoidal, sawtooth, symmetric eclipsing binary and eccentric eclipsing binary. We apply GRAPE and a BGLS periodogram to this data and show that the performance of GRAPE is superior to the periodogram on sinusoidal and sawtooth light curves with relative hit rate improvement of 18.2\% and 6.4\% respectively. The symmetric and eccentric eclipsing binary light curves have similar performance on both methods. We show the Skycam cadence is sufficient to correctly estimate the period for all of the sinusoidal shape light curves although this degrades with increased non-sinusoidal shape with sawtooth, symmetric binary and eccentric binary light curves down by 20\%, 30\% and 35\% respectively. The runtime of GRAPE demonstrates that light curves with more than 500-1000 data points achieve similar performance in less computing time. The GRAPE performance can be matched by a frequency spectrum with an oversampled fine-tuning grid at the cost of almost doubling the runtime. Finally, we propose improvements which will extend this method to the detection of quasi-periodic signals and the use of multiband light curves.
\end{abstract}





\section{Introduction}

The era of wide-field large sky surveys has facilitated the collection of data on a huge quantity of periodically variable sources \citep{b1,b2,b3,b4}. The next generation of surveys such as the Large Synoptic Survey Telescope (LSST) will further expand this catalogue of objects \citep{b5,b6}. Periodic variables include many varieties of stellar systems with different evolutionary states and morphologies. Examples of these objects include pulsating giant stars and eclipsing binary systems of many configurations \citep{b7,b8,b9}. These objects grant us insight into stellar evolution and the wider galactic evolution around the sources. The rapid classification of newly discovered candidates of these objects for follow-up is of great importance. Machine Learning techniques have been used to great effect to automate this task which is necessary given the quantity of survey data being produced \citep{b10,b11,b12,b13}.

The individual observations of a source in a given filter can be used to generate a light curve, a time-series detailing the magnitude or flux of an object at a set of given time instants. Whilst a large number of parameters have been, and continue to be, developed to characterise the variability in light curves, the three globally dominant ones are \textit{period}, \textit{colour} and \textit{amplitude} \citep{b11,b12}. Colour can be evaluated using multiband photometry with many surveys collecting data in at least three filters. Period and amplitude have a more complicated interdependency with the amplitude being a function of a candidate period with a variability class dependent relationship \citep{b10,b11}.

It is therefore clear that the successful extraction of the period for a periodic source is of great importance to successful and reliable object classification. Many periodic variable types such as the Mira-type red giant long-period variables, the classical cepheid giants and the population II cepheid giants inhabit a strict range of periods \citep{b14,b15,b16,b17}. This is due to the properties of the stellar atmosphere pulsations driving the variability of these objects \citep{b18}. Mira-type stars exhibit strong, multi-hundred day long nearly sinusoidal variations with amplitudes of greater than $2.5$\,mag in the optical bands \citep{b14}. The classical cepheids have shorter periods of days to a few months and smaller amplitudes of a few tenths to about $2$\,mag in the optical bands, usually positively correlated with the period \citep{b15}. The shape of these variations is not sinusoidal and is closer to an asymmetrical sawtooth \citep{b16,b17}. This can result in difficulties for methods designed to detect sinusoidal variability. Some periodic sources such as eclipsing binaries have a large number of configurations with a wide range of periods \citep{b19}. In these cases the precise determination of the period of these eclipses is still required for the generation of the characteristic eclipse shapes \citep{b20}.

In this paper we introduce a new approach to facilitate the rapid estimation of light curve periodicity through the use of a tuned genetic algorithm named GRAPE: Genetic Routine for Astronomical Period Estimation. This method combines genetic parameter optimisation with astronomical period estimation functions and alias correction routines allowing the quick and accurate evaluation of candidate periods across a large potential period range for light curves with many observations \citep{b21}. The fitness function for a given candidate period is determined by a Bayesian Generalised Lomb-Scargle (BGLS) periodogram capable of detecting periodic variability in light curves with substantial noise components \citep{b22,b23}. Genetic algorithms, by the virtue of their ability to quickly move around large, high dimension, parameter spaces allow this period estimation task to be performed with a high speed even on light curves with thousands to tens of thousands of data points. The versatility and speed of this method is offset by a difficulty in fine-tuning the solution as genetic algorithms sacrifice absolute accuracy on a final solution with finding it in such a large parameter space.

In \textsection2 we briefly review the different metrics possible for characterising variability at a candidate period and discuss the difficulties inherent to traditional frequency spectrum approaches. In \textsection3 we introduce GRAPE and describe how the genetic algorithm is able to optimise across the period space to the best performing candidate period based on the chosen fitness function. We also discuss the tuning parameters such an approach requires as well as the properties of 'white noise jitter' an important argument in the BGLS method \citep{b22}. In \textsection4 we evaluate the peformance and runtime of GRAPE in the detection of periodicity in synthetic light curves of different shapes. We also compare the performance with two different sampling cadences, a more traditional survey cadence and a variable cadence generated from time instants extracted from SkycamT, a wide-field camera mounted on the Liverpool Telescope, images \citep{b24}. We discuss our results and conclude with the possibilities of this method and the future work which will expand its capability in \textsection5.

\section{Period Estimation Measures}
\subsection{Periodograms}
The identification of periodic components in astrophysical signals is usually accomplished through a decomposition of the signal using a Periodogram \citep{b25,b26}. Periodograms transform a time-series into frequency space using techniques like the Discrete-time Fourier Transform (DFT) and associated Fast Fourier Transform (FFT) \citep{b27}. These powerful techniques are of limited use in astronomy due to the uneven sampling common to observational science. As a result, an alternative metric based on Fourier transforms was proposed by two independent approaches and is known as the Lomb-Scargle periodogram (LSP) \citep{b28,b29}.

When viewed from a maximum likelihood viewpoint, this method computes sinusoidal models across candidate frequencies and determines the resulting best fit \citep{b33}. Fluctuating phase offsets for differing candidate frequencies are calibrated by a phase correction term in the LSP calculation. The statistic, often referred to as spectral power, is computed for a set of candidate frequencies linearly distributed with a given density named the frequency spectrum. The method is capable of operating on unevenly distributed measurements allowing direct application to observational data. Unfortunately, it suffers from a number of drawbacks. The data points are not weighted in the calculation allowing photometrically poor data to unduly influence the quality of the fits. The metric does not provide for a constant offset requiring the data points to be centred on zero. This is commonly performed by computing the mean/median of the data points and subtracting this value. Whilst this approach is often satisfactory, if the sample mean/median is different to the underlying population mean/median, it will result in erroneous frequencies. These two drawbacks have been addressed with the computation of a more sophisticated Periodogram named the generalised Lomb-Scargle periodogram (GLS) \citep{b30}. As the underlying model is sinusoidal the LSP and GLS both exhibit decreased performance on non-sinusoidal astrophysical signals such as sawtooths and eclipsing binaries \citep{b13,b31}. Introducing additional harmonics for the sinusoidal model was proposed to allow for better fits to these non-sinousoidal signals \citep{b32}. This was found to also introduce ambiguity into the period due to increased response in fits at periods which are multiples or submultiples of the real period of the signal \citep{b33}.

The spectral power statistic is a somewhat abstract measurement but by introducing Bayesian symbolism to the best-fit statistic, the probability of a sinusoidal fit at a candidate frequency is present in the light curve given a sinusoidal prior resulting in a Bayesian Generalised Lomb-Scargle Periodogram (BGLS) \citep{b22,b23}. This takes into account the error associated with the photometric measurements although it also requires an additional term named white noise jitter. This term is required to describe the amplitude of other possible underlying signals and correlated noise in the data and underestimating this value will result in the probabilities for all candidate periods collapsing to zero with a total loss of usable information.

\subsection{The frequency spectrum}

All the above methods produce a statistic which is a function of period or frequency. Of course, this makes perfect sense as determining strong periods is the result we are hoping to obtain. However, all these approaches select trial periods through the creation of a frequency spectrum. This discretised vector of candidate periods limits the precision of best-fit period the algorithms can present. Whether the method looks for a minimum or a maximum of a given statistic, or perhaps something more complicated, the period associated with this optimised candidate period is still one of these initial candidates. Period estimation pipelines often use a fine-grid search to mitigate this by performing ever finer grids around strong candidates which results in increased computational effort \citep{b10,b12,b13}. It is also likely that many of the trial frequencies contain nothing of value and large sections of the frequency spectrum may be filtered out prior to more computationally expensive calculations. Yet, despite the capability of frequency spectrum approaches, it is clear that the true period parameter space is a continuous variable space. The discretised frequency spectrum can be oversampled to a degree where frequency estimation errors associated with the finite observational data are larger than the gaps between consecutive candidate frequencies. As true frequencies produce gaussian shape peaks due to finite-length data \citep{b33}, a candidate frequency which is slightly out of position from the true frequency might only be evaluating the side of the gaussian peak. This would result in a lower valued statistic than would be appropriate for the true period resulting in an increased potential for an incorrect result. This can be seen in long period variability candidates as the frequency spectrum is reciprocal to the period space. The treating of the period parameter space as a continuous variable allows for the correction of this artefact.

Trial period extraction using the bands method was used as a followup to the development of the Correntropy Kernalised Periodogram (CKP) \citep{b41}. Due to the increased computational complexity of this calculation, it was determined that preselecting trial periods was optimal \citep{b42}. The bands method was introduced as a method to conduct this preselection. The approach operates on the idea that data points of similar magnitude are likely to be found at integer multiples of any underlying period within magnitude percentile bands which contain a strong signal component. A candidate variable light curve can be dismantled into a number of percentile bands. $N_b$ percentile bands have their spectral window function computed on a linearly spaced frequency grid and the resulting $N_t$ local maxima are retained. This results in $N_b N_t$ trial periods for further analysis with the values of $N_b$ and $N_t$ chosen as a trade off between computational load and correct trial estimation \citep{b42}. This trial period selection approach, whilst still relying on frequency grids, suggests that it is possible to rapidly evaluate sections of the candidate period parameter space. With our proposed method, we consider an approach that can treat the period parameter as a continuous variable whilst simultaneously exploring the period parameter space and isolating regions of interest to optimise to the true underlying period.

\section{Period Estimation with GRAPE}
\subsection{Genetic Algorithms for period estimation}
Period estimation tasks are best described as a global optimisation problem across a continuous parameter space. This optimisation must be solvable with a minimal computation of trial periods to identify a global minimum in a cost function. This is a function of the parameter we wish to evaluate, in this case the period of a light curve, and the trial period at the global minimum is the desired result \citep{b21,b43}. Many optimisation routines have been developed but the complexity of the underlying cost function can rapidly diminish the performance of many problems until they are no better than a brute force approach. Evolutionary algorithms are inspired by biological evolution and are capable of exploring a large, possibly high dimesion, parameter space efficiently whilst also selecting good candidate results with a high precision \citep{b21}. Genetic algorithms are the most popular subset of evolutionary algorithms and utilise computational variants of many well known staples of biological evolution \citep{b44}. These include natural selection (survival of the fittest), genetic recombination, inheritance and mutation \citep{b43}. Genetic algorithms have been utilised in a number of problems in astrophysics \citep{b43} including period estimation \citep{b21}. To our knowledge, we have not seen them employed with current generation periodograms using their highly capable models in the form of a fitness function whilst allowing the genetic algorithm to optimise to the desired period result without requiring a frequency spectrum.

\begin{figure}
 \includegraphics[width=\columnwidth]{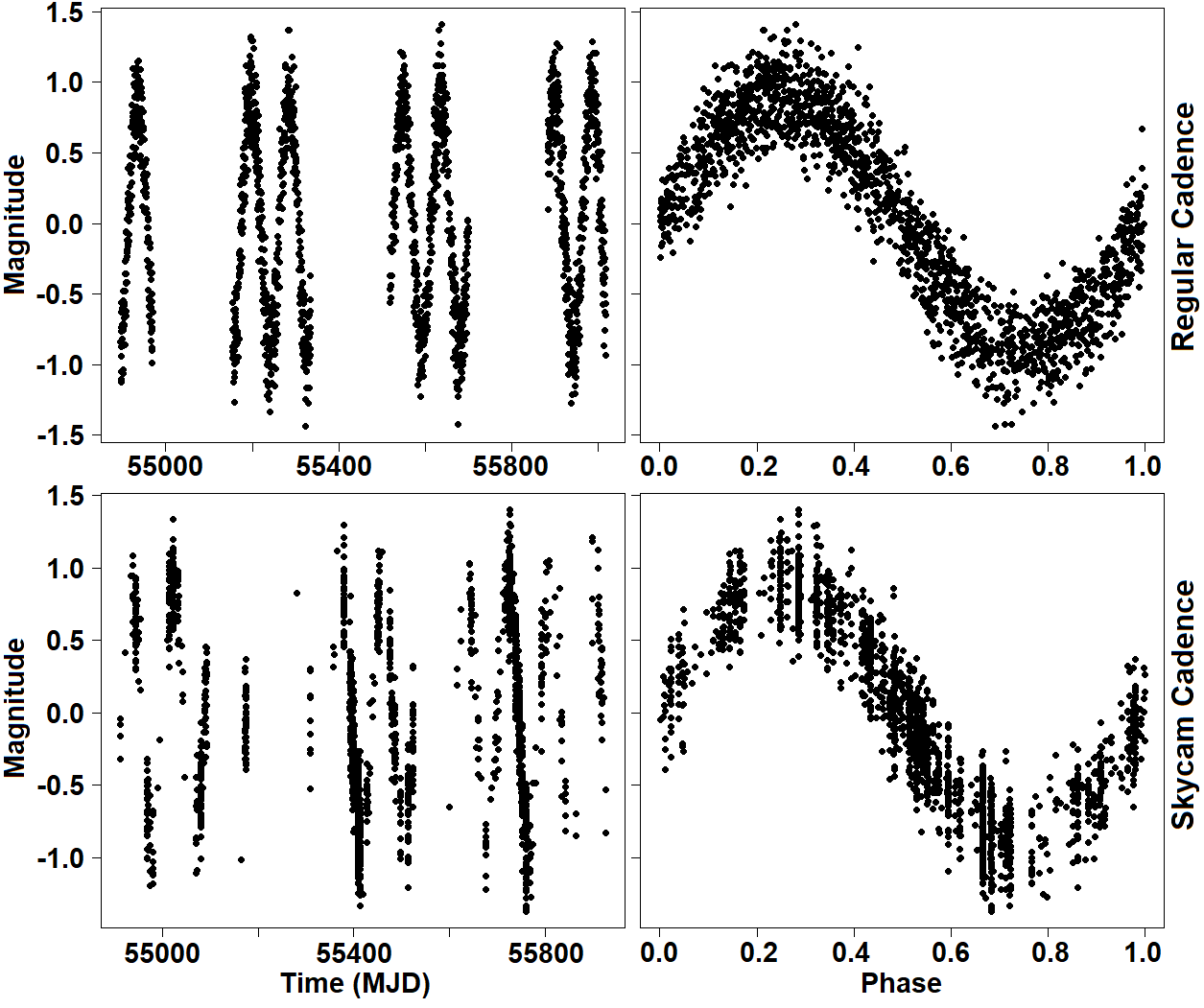}
 \caption{A simulated sinusoidal signal with white noise sampled with more traditional regular cadence and the Skycam highly variable cadence. Top left is the regular cadence raw light curve, top right is the regular cadence phased light curve, bottom left is the Skycam cadence raw light curve and bottom right is the Skycam cadence phased light curve. Skycam cadence clearly introduces new structures into the data which can produce spurious and aliased results in period estimation tasks.}
 \label{fig:cadence}
\end{figure}
To explain how these processes are inherent in our approach, we must describe how a genetic algorithm can be constructed for the period estimation task. We introduce the method we developed named GRAPE: Genetic Routine for Astronomical Period Estimation. The method was needed to improve the performance of period estimation on the Skycam database, a collection of light curves with a three year baseline produced by the reduction of images from SkycamT, a wide field optical instrument on the Liverpool Telescope \citep{b51}. This instrument co-points with the telescope and takes 10 second exposures every minute the telescope is in operation with a 35 second readout time \citep{b24}. The instrument, at the time of the collection of these images between March 2009 and Match 2012, had a Field of View (FoV) of $21^{\circ}\times21^{\circ}$ and, whilst unfiltered, had a limiting magnitude which was USNO-B catalogue calibrated to approximately 12\,mag in the R band \citep{b24,b45}. As a result of not controlling the motion of the telescope, Skycam has a unique variable cadence compared to more regular surveys. There are no guarantees when an object will be resampled and often a cluster of data points with a sampling of minutes are seperated by gaps many days long. This can lead to difficulties such as aliasing where sampling periods reflect signals across the parameter space. Figure \ref{fig:cadence} demonstrates the difference between Skycam cadence and regular cadence for a simulated sinusoidal signal with white noise.

\subsection{GRAPE initialisation method}
Genetic algorithms operate by employing a set number of individuals (or phenotypes) named the population. These individuals are a set of candidate solutions (in our case candidate periods). The individuals are distributed across the feature space in either a random or in an organised approach, such as from a previously identified underlying distribution. Despite these varying degrees of complication, generally the improvement from a uniformly random distribution is minimal. We describe how GRAPE establishes its initial population below as it requires definitions of the parameter space. This population is evolved from generation $i$ to generation $i+1$ until a cut-off point has been reached, either when the best answer reaches a given precision or a predefined number of generations has been computed \citep{b21}.

The parameters of the solution, which in our case is just a single period parameter, must be encoded into a string named a chromosome with each character named a gene. GRAPE utilises base-10 chromosomes where each gene can have an integer value $I \in \mathbb{Z} \in [0,9]$ where $\mathbb{Z}$ is the set of integers. Our encoding process is very simple and is dependent on the size of the period parameter space to be explored. GRAPE operates in frequency space and the frequency search is performed from a minimum frequency $f_{\mathrm{min}}$ defined in Equation \ref{eq:fmin}.
\begin{equation}
f_{\mathrm{min}} = \frac{1}{t_{\mathrm{max}} - t_{\mathrm{min}}}
\label{eq:fmin}
\end{equation}
where $t_{\mathrm{max}}$ is the time instant of the last data point and $t_{\mathrm{min}}$ is the time instant of the first data point to a maximum frequency of $f_{\mathrm{max}} = 20$ \citep{b45}. All units of frequency are in cycles per day\,$(d^{-1})$. For any candidate frequency $f_{\mathrm{can}}$, which always obeys $f_{\mathrm{min}} \leqslant f_{\mathrm{can}} < f_{\mathrm{max}}$, the encoded chromosome is generated by first rescaling the frequency space so that any candidate frequency $\in [0,1]$ using Equation \ref{eq:fscaled}.
\begin{equation}
f_{\mathrm{scaled}} = \frac{f_{\mathrm{can}} - f_{\mathrm{min}}}{f_{\mathrm{max}} - f_{\mathrm{min}}}
\label{eq:fscaled}
\end{equation}
The frequency is rounded to $10$ decimal places and the value of each decimal is encoded into the associated gene creating chromosomes containing 10 genes. Using this calculation and its inverse the genetic algorithm can perform generational updates on the chromosomes and extract new candidate frequencies for testing. Whilst we discuss our treatment of period as a continuous variable, it is important to recognise that we still have a precision limit with this method. The base-10 chromosomes can encode ten decimal places of the normalised period space. Skycam light curves have a baseline of approximately 1000\,days which results in a precision of $10^{-10}$\,days for extremely short periods and $10^{-3}$\,days for periods close to the maximum period. Light curves with baselines of up to several hundred years should still maintain a $0.1$\,day precision at candidates close to this maximum. A frequency spectrum from 0.05\,days to 1000\,days and an oversampling factor of 10 would only achieve equivalent worst-case precision on periods under 3\,days, and are as low precision as 2+\,day precision above 100\,days. Therefore, we believe that our precision is sufficient to justify our description of a continuous parameter space.

\begin{figure}
 \includegraphics[width=\columnwidth]{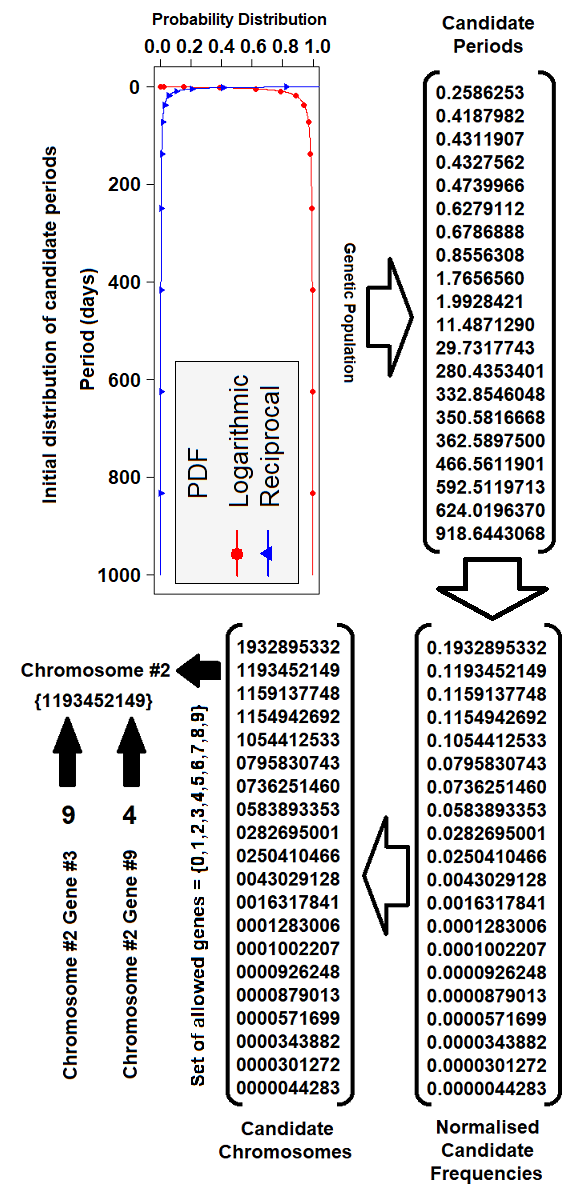}
 \caption{A demonstration of the creation of the initial population of candidate frequency chromosomes. The top left plot shows the probability distributions used to generate the randomly initialised candidates. The red line with circular points is the logarithmic distribution from equation \ref{eq:pop1} and the blue line with triangular points is the reciprocal distribution from equation \ref{eq:pop2}. The top right set shows 20 candidate periods drawn from these two distributions, 10 from each distribution. The bottom right set shows these periods are transformed into frequencies and normalised based on the minimum and maximum allowed frequencies for the optimisation. In the bottom left set the candidate frequencies are rounded to 10 decimal places and these 10 decimals become the 10 genes which makes up the genetic chromosome for each candidate. The chromosomes are strings of ten digits where each digit represents a gene. This representation allows precision with our $0.05$ to $1000$ day period range of $10^{-3}$\,days at the long period extreme to $10^{-10}$\,days at the short period extreme. This precision allows the optimisation to treat the parameter space as a continous variable.}
 \label{fig:pdf}
\end{figure}

The initial GRAPE population (individuals of generation $1$) of number $N_{\mathrm{pop}}$ is generated using two distributions to sample sufficiently across the parameter space prioritising regions where periodic variability is common. The first half of the population is generated using a distribution which is base-10 logarithmic in frequency space shown in Equation \ref{eq:pop1}.
\begin{equation}
\left\{\mathrm{Pop}_1\right\}_i = 10^{\mathrm{\textit{runif}}\left(\log_{10}(f_{\mathrm{max}}) - \log_{10}(f_{\mathrm{min}})\right) + \log_{10}(f_{\mathrm{min}})}
\label{eq:pop1}
\end{equation}
where $\left\{\mathrm{Pop}_1\right\}_i$ is the $i^{th}$ first half set candidate frequency, rerun for $i = 1, ..., 1000$ to give $\left\{\mathrm{Pop}_1\right\}$, the set of first half set candidate frequencies, \textit{runif} is a set of uniformly distributed random numbers generated between $0$ and $1$ and $f_{\mathrm{min}}$ and $f_{\mathrm{max}}$ are as above. This distribution skews the population towards lower frequencies and therefore higher periods yet it still heavily samples the low period end of the parameter space as it is a function of frequency. The second half of the population is produced by a function linear in period space and therefore a reciprocal in frequency space. This set of individuals is generated using Equation \ref{eq:pop2} with the same definitions as above.
\begin{equation}
\left\{\mathrm{Pop}_2\right\}_i = \frac{1}{\mathrm{\textit{runif}}\left(\frac{1}{f_{\mathrm{min}}} - \frac{1}{f_{\mathrm{max}}}\right) + \frac{1}{f_{\mathrm{max}}}}
\label{eq:pop2}
\end{equation}
where $\left\{\mathrm{Pop}_2\right\}_i$is the $i^{th}$ second half set candidate frequency, rerun for $i = 1, ..., 1000$ to give $\left\{\mathrm{Pop}_2\right\}$, the set of second half set candidate frequencies. This distribution is highly skewed to low frequencies and therefore high periods and is required to ensure the long period end of the parameter space is sufficiently explored. Had we used a distribution that was simply linear in frequency space it would often remain in the low period end of the parameter space resulting in a heavy loss of performance. Figure \ref{fig:pdf} demonstrates this candidate generation process showing a set of generated candidate periods.
The candidate frequencies must then be sorted based on their performance on a fitness function which computes how well they fit the data. This sorting is then used to determine which candidate frequencies should be retained to subsequent generations. GRAPE uses the Bayesian generalised Lomb-Scargle periodogram (BGLS) as its fitness function \citep{b22,b23}. This method is quickly computable due to a lack of any operation more complex than simple summations. Despite this light computational complexity, it was one of the best performing periodograms when trialed on SkycamT light curves which had been matched to known periodic variable stars through cross-matching with the American Association of Variable Star Observers (AAVSO) Variable Star Index (VSI) catalogue. We surmise this is likely due to the methods flexibility on population mean and data point weighting combined with the ability to control the signal and noise components through the careful use of the white noise jitter argument. Whilst we make use of BGLS as our fitness function, it is important to note that the genetic algorithm design is highly modular and other periodograms or combinations of periodograms may be used. Our main priority was to maintain high accuracy across the entire period space whilst limiting the processing time on the Skycam data which exhibits 136,420 light curves with more than 2000 data points which can be computationally intensive with the frequency spectrum approach \citep{b24}.

\subsection{Evolutionary implementation}
The candidate frequencies chromosomes, sorted by the BGLS fitness function, must be used to create the next generation propagating through knowledge gained from the initial population whilst allowing flexibility to explore regions that were not initially scanned. Genetic algorithms use a mechanism analogous to sexual reproduction to generate the subsequent generation. A number of the previous generation are paired up into $N_{\mathrm{pairups}}$ partners. Unlike sexual reproduction, the pairups can be with two copies of the same individual if the genetic algorithm selects it. It is also important to utilise a parameter named \textit{selection pressure} for this operation. This determines what quantity of the fittest individuals are selected to reproduce, the parents. Selection pressure can take several forms such as Tournament selection where random subsets of the individuals compete to be selected as a parent or Roulette selection where a probability of selection is assigned to each individual summing to one based on their fitness function evaluation. For GRAPE we have only tested the \textit{Roulette} selection pressure method which is superior at preserving important genetic diversity than simply selecting the top best-fitting individuals. We employ an argument named $P_{\mathrm{fdif}} \in \mathbb{R} \in [0,1]$ which defines the contribution of the sorted fitness function to selecting the parent individuals. The argument $P_{\mathrm{fdif}} = 1$ results in a high contribution and essentially results in the best performing individual pairing up with itself $N_{\mathrm{pairups}}$ times. Whilst it would seem to be a good idea as it uses the best fitting frequency, the loss of all other genetic information greatly hinders the algorithms ability to explore other areas of the frequency space. The algorithm would likely get stuck at one of the intial candidate frequencies and fail to find the true frequency. On the otherhand, $P_{\mathrm{fdif}} = 0$ means that the parents are chosen randomly from the previous generation meaning very little learned knowledge of the fitness function is propogated to the next. This results in an algorithm that haphazardly jumps around the parameter space randomly requiring a 'lucky hit' in one of the final generations to produce a reasonable result. Ultimately, a value must be determined that results in a propogation of fitness information without a complete loss of genetic diversity. It is also extremely important that the rank number for the sorted chromosomes be used, not the raw BGLS statistic, or else the selection might still focus on the best performing chromosome despite the selection pressure. This is due to the large difference between BGLS statistic values for even two similarly performing candidate frequencies \citep{b21}. Selection pressure is what introduces the survival-of-the-fittest into our approach.

\begin{figure}
 \includegraphics[width=\columnwidth]{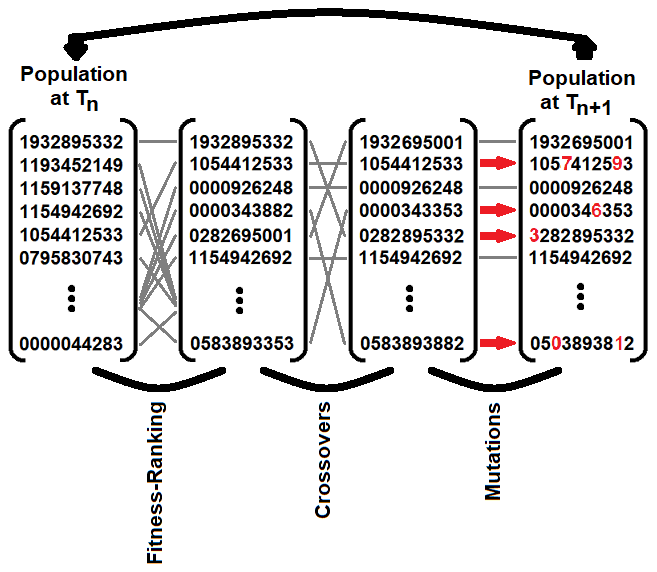}
 \caption{A demonstation of the primary evolutionary methods utilised by GRAPE. As seen in figure \ref{fig:pdf}, the candidate frequencies are expressed as a set of chromosomes encoding the frequency information. A set of the population is selected to produce the next generation. The first operation utilises the fitness function, in our case a BGLS function. This function is used to rank the candidate frequencies in order of their response as the candidates selected for reproductive operations are dependent on their model performance. The crossover operation selects two chromosomes and generates two children by splitting the chromosome at a given point between two genes and placing the starting genes of parent one with the final genes of parent two and vice versa. Finally, the mutation operation can select any of the genes for a change, also known as a `copy error'. Once complete, the newly generated chromosomes are ranked and become part of the next generation. This generational update is then repeated as many times as required to produce a population of high performing individuals based on the chosen fitness function.}
 \label{fig:ga}
\end{figure}

\begin{figure}
 \includegraphics[width=\columnwidth]{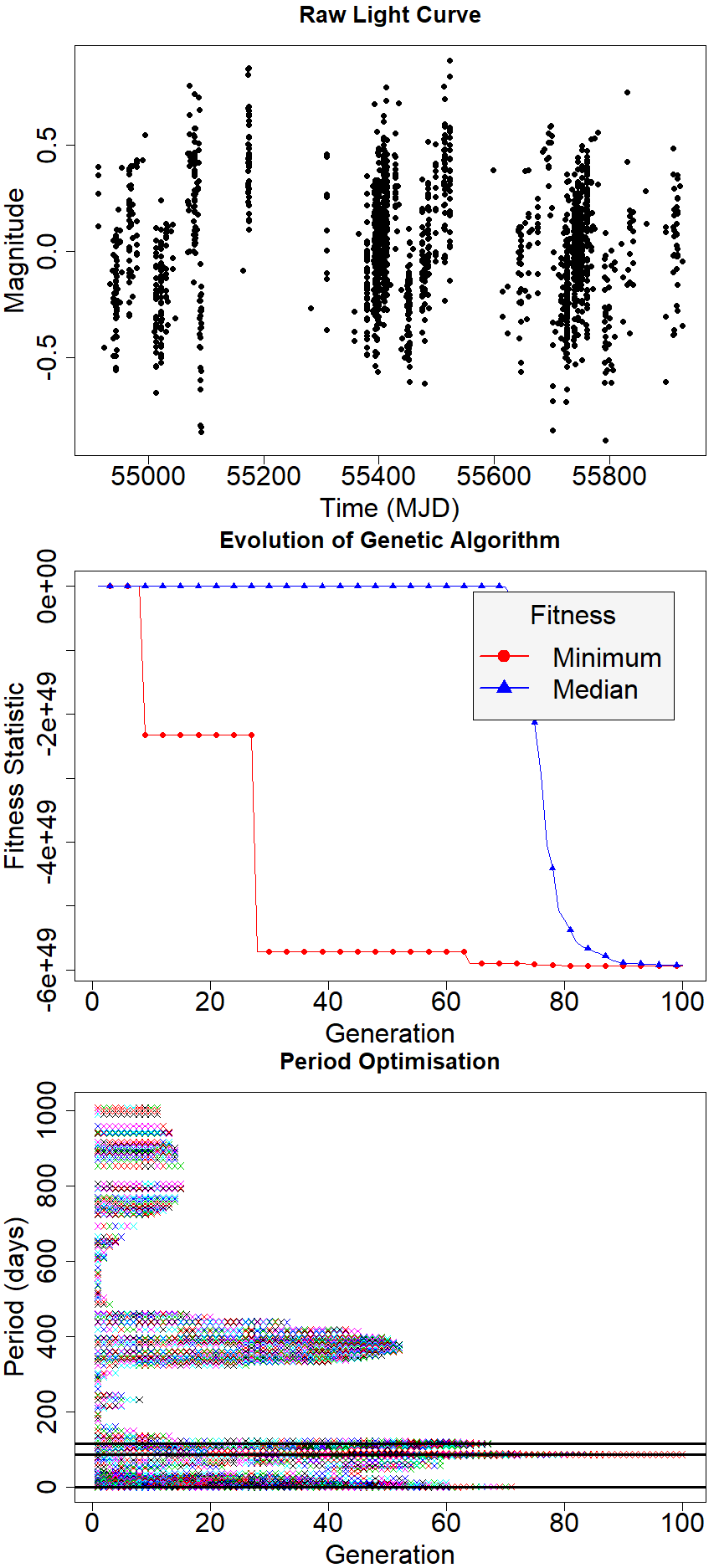}
 \caption{Plots of the exploration of the period space against generation number for a simulated sawtooth light curve. The top plot shows the simulated light curve that GRAPE is processing. It is a Skycam cadence simulated sawtooth with a period of 87.58\,days and a signal-to-noise ratio (SNR) of 2. It has 1721 data points sampled in yearly seasonal windows with a strong diurnal sampling. The middle and bottom plots demonstrate the evolution of the GRAPE genetic algorithm across the 100 generations of genetic optimisation. The middle plot shows the minimum and median statistics of the BGLS fitness function for every candidate in each generation. The red line with circular points is the minimised candidate for each generation and the blue line with triangular points is the median of the fitness of every candidate for each generation. The bottom plot shows the location of each generations candidate frequencies in period space. The candidate periods selected by the genetic optimisation are demonstrated by the black horizontal lines.}
 \label{fig:posit}
\end{figure}

Upon the selection of the $N_{\mathrm{pairups}}$ reproduction events, the children of these events must be determined. Each two parents produce two children as part of the reproduction. These children inherit genetic information from their parents but may also be a new unique formulation of this information. This means that the reproduction will often produce two new candidate frequencies for evaluation in the next generation. These frequencies use genetic information from the previous generation whilst still exploring a new part of the parameter space. The two mechanisms used for this modification of the parents chromosomes are called crossover and mutation. Crossover is analogous to how children inherit traits from both their parents by splitting the parent chromosomes into sections and then recombining them into two children with combinations of both parents chromosomes. GRAPE utilises the simplest form of crossover by performing a single split on both parent chromosomes. First, an argument named $P_{\mathrm{crossover}} \in \mathbb{R} \in [0,1]$ determines the probability that any given reproductive event will result in a crossover. If a crossover is triggered during a reproduction, a uniform random number generator selects an integer value $\in \mathbb{Z} \in[1,10]$. This determines at what location the split will occur. Child chromosome 1 will contain the genes from parent chromosome 1 up to the split location and the genes from the parent 2 chromosome after the split location. Child chromosome 2 is the inverse with the first part being from parent 2 and the second part from parent 1. If no crossover occurs, the children chromosomes are identical to the parent chromosomes at this point. Next the mutations are calculated. Mutations are randomly selected gene changes inside the children chromosomes. For each gene in the two children chromosomes, an argument \textit{mutation} determines the probability that a mutation occurs. A uniform random number generator $\in \mathbb{R} \in[0,1]$ generates a value for each of the 20 genes in the two children chromosomes. Any genes for which the associated generated value is below or equal to $P_{\mathrm{mutation}}$ is determined to have undergone a mutation. For each mutated gene, a uniform random number generator computes an integer value between $0$ and $9$ and assigns this value as the new gene value at that location. Mutation allows the genetic algorithm to create new candidate frequencies in previously unvisited locations without requiring the two parents to contain this information in their chromosomes. As each child gene is subjected to this mutation probability, it is usually recommended to use a low value for the later generations to prevent the genetic algorithm from jumping away from optimised answers. Upon the completion of the reproductive events, any new child chromosomes are evaluated using the BGLS and appropriately ranked by the fitness function. Figure \ref{fig:ga} shows an example of this genetic evolution using some of the chromosomes created in figure \ref{fig:pdf}.
These operations result in a generation $i+1$ population of size $N_{\mathrm{pop}} + 2 N_{\mathrm{pairups}}$. As each subsequent generation would produce new offspring, the population size would rapidly increase introducing extra computational complexity with zero benefits. Therefore, we must remove $2 N_{\mathrm{pairups}}$ candidate frequencies from the population in a process analogous to biological death with the exception that, if the algorithm chooses it, any individual could live forever. We employ an `anti-selection' pressure to accomplish this which we call the \textit{death fraction}. This is, like selection pressure, a probability $P_{\mathrm{dfrac}} \in \mathbb{R} \in [0,1]$ which determines what proportion of the removed candidate frequencies were from the poorest performing BGLS frequencies verses randomly eliminating chromosomes of any fitness ranking. The only difference from selection pressure, is that we always retain the best performing candidate frequency from the current generation to preserve its genetic knowledge. This last step produces the individuals of generation $i+1$ and the cycle continues as described towards the production of generation $i+2$. This cycle is repeated until a predetermined cut off or generation $N_{\mathrm{gen}}$ is produced. The fittest individual from this final generation is then returned as the genetically chosen fittest chromosome. It is then decoded into a chosen frequency and returned to the user. Figure \ref{fig:posit} demonstrates how the exploration of the period space evolves across the generations due to the propagation of genetic information on the performance of the fitness function for a sinusoidal light curve. It is clear that the whole period space is investigated initially and by generation 40 the only remaining regions are the true period and its multiples. By generation 80 all regions of the period space apart from that located at the best fit period are discarded. The final generations are utilised for fine tuning the ultimate result.

\subsection{Vuong Closeness test for periodic model discrimination}
Many period estimation methods including the Bayesian Generalised Lomb-Scargle used in the GRAPE fitness function suffer from common failure modes as a result of the cadence of a light curve and non-sinusoidal shape. This translates into the periodogram as multiple significant peaks of which any one may be the true astrophysical period. Therefore our genetic algorithm must be capable of identifying not just a global optimum but clusters of persistent local optima which are then optimised to a candidate period. Additionally, whilst it would be preferred if the chosen frequency could be trusted as the best possible result for a given set of data, genetic algorithms suffer from the disadvantage that they are somewhat a black box. The propogation of the candidate frequencies from generation to generation is heavily controlled by randomness sourced from random number generators as obvious from the method described above. This can result in differences in the exploration of the parameter space purely due to the use of different random number seed values in each run. Therefore, GRAPE has been designed to detect $N_t$ trial periods through the use of k-means clustering for each generation of the routine where $k = N_t$ with a `dominant' seed value. The clusters with a standard deviation below a critical value $\sigma_t$ are recorded per generation. The routine continues until standard deviation of the cluster means for a given generation drops below $\sigma_t$. The cluster means are then rounded to 2 decimal places and analysed for repetition across multiple generations. The top $N_t-1$ cluster means are selected as candidate periods along with the global minimum of the genetic run. Due to the randomness of the genetic algorithm this operation is performed a second time with a different `submissive' seed value. In the event that any of the candidate periods are close repetitions of any other candidate period, the repeated candidate periods are replaced with new candidate periods from the submissive run.

Upon the production of $N_t$ candidate periods which may or may not be similar, they are fine tuned through the use of $N_t$ single-period genetic runs with $N_{\mathrm{finegen}}$ generations to achieve an optimised result for a period range of $\pm 10\%$ of the candidate period. The $N_t$ candidate periods are then tested to determine which period produces the best performing fit to the light curve. It is possible to use the BGLS to determine the best performing of these frequencies, however, GRAPE utilises a more powerful information-theoretic statistic. The Vuong closeness test is a statistical model similarity measure based on the Kullback-Leibler Information Criterion \citep{b46}. This method was proposed in the discrimination of aliased periods during the period estimation task \citep{b32}. Aliased periods are reflections of true periodic signals with a sampling period, also known as a spurious period. They are calculated by Equation \ref{eq:alias} \citep{b52}.
\begin{equation}
P_{j,k} = k \frac{t_{sid}}{(t_{sid}/{P_t})+j}
\label{eq:alias}
\end{equation}
where $P_{j,k}$ is a set of alias periods produced by a trial period $P_t$, $t_{sid} = 0.99726957$\,days is the sidereal day, $k$ is an integer value from a vector of values, $k = [1,2,3]$ which defines the multiples of the trial period and $j = [-3,-2,-1,-0.5,0,0.5,1,2,3]$, the set of possible alias frequencies limited to $|j| \leq 3$. This is due to higher values of $j$ producing aliases with fitness function response of similar order to the noise continuum \citep{b33}. For each of the $N_t$ periods, GRAPE generates 27 sinusoidal regression models for the 27 $P_{j,k}^{-1}$ frequencies with 4 parameters, an intercept, a linear trend and a sine and cosine component. These models are used to compute the Vuong Closeness statistic between the trial period $P_t$ model and each of the $P_{j,k}$ models. If the Vuong Closeness statistic indicates one of the $P_{j,k}$ models outperforms the trial period, the trial period is replaced with the Vuong Closeness maximising candidate period as long as it is not a known spurious period calculated previously.

Upon the generation of the chosen $N_t$ periods by the genetic algorithm, GRAPE generates another set of $N_t + 2$ sinusoidal regression models with 4 parameters, as above. The frequency of these Fourier components are defined as the $N_t$ genetically chosen frequencies, a constant model with only an intecept and a daily model with a period of one day. Finally, the $N_t$ models for the chosen periods are compared using the Vuong closeness test to select the final period. GRAPE offers two options for this method. In the first, the model of every chosen period from each rerun are compared. This requires the computation of $N_t^{2}-N_t$ Vuong closeness tests. Alternatively, if the value of $N_t$ is prohibitively high (although this many reruns is unlikely), the chosen periods are all compared to a constant model with no linear or sinusoidal terms. This requires $N_t$ Vuong closeness tests. This method does have the disadvantage that it may screen out the correct period in favour of one due to the sampling and therefore we recommend the first option. This last step completes the GRAPE routine and results in a determined periods.

The chosen period is used to compute the Vuong Closeness statistic between the chosen frequency model and the constant and one sidereal day model (a sinusoidal model with a frequency of $t_{sid}^{-1}$\,cycles/day). These statistics are used to describe the significance of the chosen period relative to a purely non-periodic model as well as comparing any periodicity detected to the one sidereal day dominant spurious period. A significant periodic signal may produce a high value against the constant model, but will have a much lower value in the one day model if it is due to a sampling periodicity. A real astrophysical signal would be expected to be significant in both of these statistics. GRAPE ulimately returns for a given light curve, an optimised period which has been checked for multiplicity and aliasing with a sinusoidal model and two Vuong closeness test statistics for the sinusoidal model of this optimised period calculated against a constant signal model and a sinusoidal signal with a period of the sidereal day. No confidence margins are applied to the Vuong closeness statistics although they can be generated from them.

\section{Evaluation of Period Estimation}
\subsection{Experimental Design}
We presume that GRAPE should exhibit improved performance over a standard frequency spectrum approach due to the treatment of period as a continuous variable. In terms of the processing time for each light curve, GRAPE should require less calculations than a frequency spectrum as only newly computed candidate frequencies must be evaluated and low interest areas of the period space can be avoided.

We designed an experiment using simulated light curves in order to compare the performance of GRAPE against the traditional BGLS, with a dense frequency spectrum, periodogram. As our method is designed as a component of a classification pipeline for the Skycam instruments, we decided to produce two sets of light curves. The first set uses the variable Skycam cadence produced by an instrument with no control over the movement of the parent telescope. The second simulates a more traditional cadence with seasonal gaps added to reproduce light curves similar to standard surveys. We started with the generation of the Skycam cadence time instants. We accessed the SkycamT database and randomly selected 1000 light curves, of which 250 had $100 \leqslant n < 200$,  250 had $200 \leqslant n < 500$, 250 had $500 \leqslant n < 1000$ and the final 250 had $1000 \leqslant n < 2000$, where $n$ is the number of data points in each light curve. This was chosen to generate simulated light curves with a wide range of entries but with a median much closer to low values, a statistical trait of the Skycam survey. The time instants of these light curves recorded in Modified Julian Date (MJD) are recorded as the Skycam-cadence set. To generate the regular-cadence set we took the minimum and maximum time instant value for each light curve in the Skycam-cadence set and generated a linear grid of time instants with a separation of 0.1\,days for each light curve. This grid of time instants was phased with a period of 365.25\,days and only time instants with an associated phase of $0.0 \leqslant ph < 0.5$ are retained. Finally, $n$ of these time instants are randomly selected where $n$ is the number of data points in the Skycam-cadence time instants for each of the 1000 light curves. We then generated $1000$ periods for these two sets of $1000$ time instant vectors. We used a uniform random number generator with the function shown in Equation \ref{eq:pi}.
\begin{equation}
P_i = 10^{\mathrm{\textit{runif}} \times \left(\log_{10}(P_{\mathrm{max}}) - \log_{10}(P_{\mathrm{min}})\right) + \log_{10}(P_{\mathrm{min}})}
\label{eq:pi}
\end{equation}
where $P_i$ is test period $i$, rerun $1000$ times for $i = 1, ..., 1000$ to obtain $\left\{P\right\}$, the set of test periods, $\mathrm{\textit{runif}} \in \mathbb{R} \in [0,1]$ is a uniform random number generator, $P_{min} = 0.05$, $P_{max} = 1000$ are minimum and maximum periods in the period space we wish to optimise. We use a logarithmic projection to skew the period distribution to low periods of which there are more known object classes \citep{b10,b12,b13}. With the two lists of $1000$ light curve time instant vectors and $1000$ simulated periods, we may now generate light curves of various shapes to test our method.

\begin{figure}
 \includegraphics[width=\columnwidth]{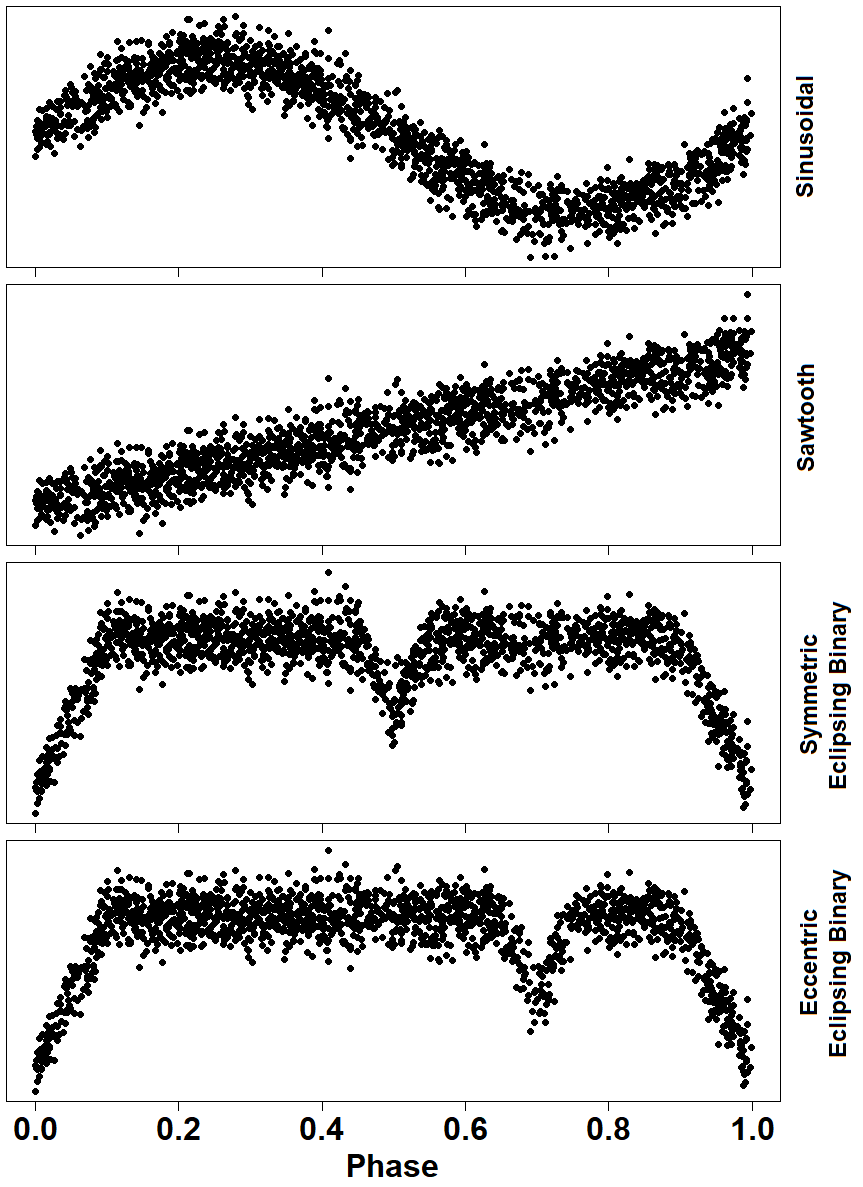}
 \caption{Phased simulated light curves of a sinusoidal, sawtooth, symmetric eclipsing binary and eccentric eclipsing binary.}
 \label{fig:lcshapes}
\end{figure}

We chose to generate light curves of four different shapes, sinusoidal, sawtooth, symmetric eclipsing binary and eccentric eclipsing binary. This resulted in $2000$ light curves of each shape, one with Skycam cadence and one with regular cadence resulting in $8000$ total light curves to test with the two algorithms. The light curves are populated with gaussian white noise using Equation \ref{eq:as}.
\begin{equation}
A_{s} = \sigma_{n} \sqrt{2 (\mathrm{SNR})}
\label{eq:as}
\end{equation}
where $A_{s}$ is the sinusoidal amplitude of the synthetic signal, $\sigma_{n}$ is the standard deviation of the white noise, and $\mathrm{SNR}$ is the desired Signal-to-Noise ratio. For this experiment all the light curves are generated with a SNR of 2 which is determined as a 0.4\,unit amplitude to a 0.2\,unit standard deviation of white noise. The sinusoidal light curves were generated using the time instants and the associated simulated period using Equation \ref{eq:sinlc}.
\begin{equation}
y_{i} = A_{s} sin\left(\frac{2\pi t_{i}}{P}\right) + \sigma_{n}\epsilon_{i}
\label{eq:sinlc}
\end{equation}
where $y_{i}$ is the magnitude value of data point $i$, $t_{i}$ is the time instant of data point $i$, $\epsilon_{i}$ is a normal distributed error value for data point $i$ with a mean of $0$ and a standard deviation of $1$ and $P$ is the simulated period. $A_{s}$ and $\sigma_{n}$ are as above. Sawtooth light curves use the function shown in Equation \ref{eq:sawlc}.
\begin{equation}
y_{i} = 2A_{s} \left(\frac{t_{i}}{P} - \left\lfloor\frac{t_{i}}{P}\right\rfloor\right) + \sigma_{n}\epsilon_{i}
\label{eq:sawlc}
\end{equation}
where $\left\lfloor x \right\rfloor$ is the closest integer to $x$ rounded down, the `floor' of $x$. For the eclipsing binary light curves, we decided to keep transit duration and shape at constant phases, i.e. they are a linear function of the underlying period. We concede this is not a perfectly physical representation as the parameters of two binary stars and their orbital properties allow many different possible eclipse shapes \citep{b19,b20} however, we wished to test GRAPE on eclipse light curve shape independent of period. Research has been conducted into the performance of eclipsing binary detection with alternate eclipse shapes and periods \citep{b48,b49}. We follow a similar process to generate both symmetric and eccentric eclipsing binary simulated light curves. First we populate the light curve with a constant signal with white noise using Equation \ref{eq:blanklc}.
\begin{equation}
y_{i} = \sigma_{n}\epsilon_{i}
\label{eq:blanklc}
\end{equation}
We then phase the simulated data points around the period. Data points between the phases $0.0 \leqslant ph_{i} \leqslant 0.1$ and $0.9 \leqslant ph_{i} \leqslant 1.0$ are located within the primary eclipse and have a triangular subtraction applied with depth of $2A_{s}$ and of base length phase of $0.2$. After these operations, the symmetric and eccentric light curve method diverges. The symmetric light curves have a secondary eclipse located at $ph = 0.5$ with a triangular subtraction of depth $A_{s}$ and a base length phase of $0.1$. The eccentric light curves have a secondary eclipse of identical size but centred at $ph = 0.7$. Figure \ref{fig:lcshapes} demonstrates the phased light curves of these four signal shapes.

The results of GRAPE and the BGLS are determined by taking the input period and the estimated period and computing if the relationship is one of six possible types: a hit, a multiple, a submultiple, a one-day alias, a half-day alias or an unknown mode. A hit is when the estimated period matches the estimated period to within a tolerance and is true if it satisifes the inequality shown in Equation \ref{eq:hit}.
\begin{equation}
|P_{i} - P_{e}| < \epsilon P_{i}
\label{eq:hit}
\end{equation}
where $P_{i}$ is the input period, $P_{e}$ is the estimated period and $\epsilon$ is the tolerance. A multiple is a realistic integer multiple of the input period and is defined as any relationship which does not satisfy the hit inequality, satisfies $P_{e} > P_{i}$, and satisifies the inequalities in either Equation \ref{eq:mult1} or \ref{eq:mult2}.
\begin{equation}
\left\lfloor\frac{P_{e}}{P_{i}}\right\rfloor \leqslant 3 \ \ \ \ \ \ \mathrm{and} \ \ \ \ \ \ \left|\frac{P_{e}}{P_{i}} - \left\lfloor\frac{P_{e}}{P_{i}}\right\rfloor\right| < \epsilon
\label{eq:mult1}
\end{equation}
\begin{equation}
\left\lceil\frac{P_{e}}{P_{i}}\right\rceil \leqslant 4 \ \ \ \ \ \ \mathrm{and} \ \ \ \ \ \ \left|\frac{P_{e}}{P_{i}} - \left\lceil\frac{P_{e}}{P_{i}}\right\rceil\right| < \epsilon
\label{eq:mult2}
\end{equation}
where $\left\lceil x \right\rceil$ is the closest integer to $x$ rounded up, the `ceiling' of $x$. A submultiple is similar to the multiple and is a realistic integer division of the input period and is defined as any relationship which does not satisfy the hit inequality, satisfies $P_{e} < P_{i}$, and satisifies the inequalities in either Equation \ref{eq:smult1} or \ref{eq:smult2}.
\begin{equation}
\left\lfloor\frac{P_{i}}{P_{e}}\right\rfloor \leqslant 3 \ \ \ \ \ \ \mathrm{and} \ \ \ \ \ \ \left|\frac{P_{i}}{P_{e}} - \left\lfloor\frac{P_{i}}{P_{e}}\right\rfloor\right| < \epsilon
\label{eq:smult1}
\end{equation}
\begin{equation}
\left\lceil\frac{P_{i}}{P_{e}}\right\rceil \leqslant 4 \ \ \ \ \ \ \mathrm{and} \ \ \ \ \ \ \left|\frac{P_{i}}{P_{e}} - \left\lceil\frac{P_{i}}{P_{e}}\right\rceil\right| < \epsilon
\label{eq:smult2}
\end{equation}
If the estimated period is the one day alias of the input period, the inequality shown in Equation \ref{eq:dayalias} is satisfied.
\begin{equation}
\left|\left|\frac{P_{i}}{1 \pm P_{i}}\right| - P_{e}\right| < \epsilon
\label{eq:dayalias}
\end{equation}
The presence of a half-day alias can be determined using a similar inequality shown in Equation \ref{eq:halfdayalias}.
\begin{equation}
\left|\left|\frac{P_{i}}{1 \pm 2P_{i}}\right| - P_{e}\right| < \epsilon
\label{eq:halfdayalias}
\end{equation}
Any light curve period estimation task where $P_{i}$ and $P_{e}$ do not satisfy any of the above inequalities are declared unknown failures.

\subsection{Period estimation performance}
GRAPE ran the data set with 3 different dominant seeds $\mathrm{Seed}_D = [1,2,3]$ to screen out poor convergence events with the Vuong closeness test with the submissive seed set to $\mathrm{Seed}_S = \mathrm{Seed}_D + 100$. The input arguments were as follows: $N_{\mathrm{pop}} = 200$, $N_{\mathrm{pairups}} = 50$, $N_{\mathrm{gen}} = 100$, $N_{\mathrm{finegen}} = 50$, $P_{\mathrm{crossover}} = 0.65$, $P_{\mathrm{mutation}} = 0.8 - 0.008i$, where $i$ is the generation, $P_{\mathrm{fdif}} = 0.6$ and $P_{\mathrm{dfrac}} = 0.7$. These values were established by a grid-search cross-validation operation on a stratified subset of 100 synthetic light curves although the sinusoidal light curves indicated a lower gradient of  $P_{\mathrm{mutation}} = 0.8 - 0.003i$ on the mutation rate. This was determined to be a result of the selection of all $N_t$ trial periods in the same period region as the true period. The number of candidate periods selected by the GRAPE genetic clustering method is $N_t = 5$ which are then analysed by the Vuong Closeness test. This only applies in situations where the signal is purely sinuosidal with Gaussian noise and was therefore rejected. The linear decay gradient on the mutation value has an important effect on exploring the parameter space as well as fine tuning the final period. We discuss more about this property in the next section.

The BGLS periodogram is performed by selecting the top $N_t$ independent peak periods with an oversampling factor, which determines the density of the frequency spectrum, of $N_{\mathrm{ofac}} = 5$. The Vuong Closeness test is then applied to these $N_t = 5$ peaks as well as their multiples and aliases in a similar operation to the one performed on the GRAPE candidate periods. The best performing period model is selected as the BGLS periodogram final period. There are also a number of shared arguments for the BGLS fitness function between both GRAPE and the periodogram. The white noise jitter, which tunes the probability response for candidate periods, $\mathrm{jit} = 0.4 \cdot A_{\mathrm{lc}}$ where $A_{\mathrm{lc}}$ is the estimated amplitude of the light curve determined by Equation \ref{eq:amplc}.
\begin{equation}
A_{\mathrm{lc}} = \frac{\left|y_{max} - y_{min}\right|}{2}
\label{eq:amplc}
\end{equation}
where $y_{min}$ and $y_{max}$ are the minimum and maximum values of the measurement unit for a given light curve. The period space is $P_{min} = 0.05\,\mathrm{days}$ to $P_{max} = (t_{max} - t_{min})\,\mathrm{days}$ where $t_{min}$ and $t_{max}$ are the minimum and maximum time instants for a given light curve. The gaussian filter for spurious period removal is left unused.

The confidence intervals of the GRAPE and BGLS periodogram performances on the synthetic light curves is determined using 100,000 bootstrapped samples from the 3000 GRAPE light curve period estimations (all 3 seeds on the 1000 synthetic light curves) and 1000 BGLS periodogram period estimations. In this bootstrapping operation the results are resampled with replacement 100,000 times and used to compute a mean performance and the $95\%$ confidence intervals through the selection of the $5^{\mathrm{th}}$ and $95^{\mathrm{th}}$ percentile performance values for the hit rate, the multiple rate (sum of the multiples and submultiples) and the aliasing rate (the sum of the one day and half day aliases).

Table \ref{tab:regcadgrape} shows the results of this experiment using GRAPE on the regular cadence four light curve types with an accuracy tolerance of $\epsilon = 0.01$, a one percent allowed error in period. Note, it is possible for the row sum of a light curve to be above $1$ as some periods can satisfy both a submultiple and an alias simultaneously and we do not presume which is the mode responsible for this error. Table \ref{tab:sktcadgrape} shows the GRAPE performance on the Skycam cadence light curves, table \ref{tab:regcadpergm} for the BGLS periodogram on the regular cadence light curves and table \ref{tab:sktcadpergm} for the BGLS periodogram on the Skycam cadence light curves.

\begin{table}
 \caption{GRAPE period estimation results on regular cadence simulated light curves with a tolerance $\epsilon = 0.01$.}
  \label{tab:regcadgrape}
 \begin{tabular}{lccc}
  \hline
  Type & Hit & Multiple & Alias\\
  \hline
  Sinusoidal & 0.831 $\pm$ 0.011 & 0.001 $\pm$ 0.001 & 0.004 $\pm$ 0.002\\
  Sawtooth & 0.741 $\pm$ 0.013 & 0.006 $\pm$ 0.002 & 0.006 $\pm$ 0.003\\
  Symmetric EB & 0.032 $\pm$ 0.005 & 0.637 $\pm$ 0.014 & 0.024 $\pm$ 0.006\\
  Eccentric EB & 0.362 $\pm$ 0.014 & 0.280 $\pm$ 0.014 & 0.011 $\pm$ 0.005\\
  \hline
 \end{tabular}
\end{table}

\begin{table}
 \caption{GRAPE period estimation results on Skycam cadence simulated light curves with a tolerance $\epsilon = 0.01$.}
  \label{tab:sktcadgrape}
 \begin{tabular}{lccc}
  \hline
  Type & Hit & Multiple & Alias\\
  \hline
  Sinusoidal & 0.824 $\pm$ 0.011 & 0.000 $\pm$ 0.001 & 0.026 $\pm$ 0.007\\
  Sawtooth & 0.527 $\pm$ 0.015 & 0.007 $\pm$ 0.003 & 0.141 $\pm$ 0.004\\
  Symmetric EB & 0.015 $\pm$ 0.004 & 0.358 $\pm$ 0.014 & 0.075 $\pm$ 0.010\\
  Eccentric EB & 0.149 $\pm$ 0.011 & 0.158 $\pm$ 0.011 & 0.112 $\pm$ 0.012\\
  \hline
 \end{tabular}
\end{table}

\begin{table}
 \caption{BGLS periodogram period estimation results on regular cadence simulated light curves with a tolerance $\epsilon = 0.01$.}
  \label{tab:regcadpergm}
 \begin{tabular}{lccc}
  \hline
  Type & Hit & Multiple & Alias\\
  \hline
  Sinusoidal & 0.703 $\pm$ 0.024 & 0.001 $\pm$ 0.002 & 0.000 $\pm$ 0.000\\
  Sawtooth & 0.696 $\pm$ 0.024 & 0.002 $\pm$ 0.003 & 0.001 $\pm$ 0.002\\
  Symmetric EB & 0.031 $\pm$ 0.009 & 0.632 $\pm$ 0.025 & 0.024 $\pm$ 0.009\\
  Eccentric EB & 0.426 $\pm$ 0.026 & 0.251 $\pm$ 0.023 & 0.012 $\pm$ 0.007\\
  \hline
 \end{tabular}
\end{table}

\begin{table}
 \caption{BGLS periodogram period estimation results on Skycam cadence simulated light curves with a tolerance $\epsilon = 0.01$.}
  \label{tab:sktcadpergm}
 \begin{tabular}{lccc}
  \hline
  Type & Hit & Multiple & Alias\\
  \hline
  Sinusoidal & 0.717 $\pm$ 0.023 & 0.001 $\pm$ 0.002 & 0.030 $\pm$ 0.009\\
  Sawtooth & 0.508 $\pm$ 0.026 & 0.010 $\pm$ 0.005 & 0.143 $\pm$ 0.021\\
  Symmetric EB & 0.021 $\pm$ 0.008 & 0.382 $\pm$ 0.025 & 0.085 $\pm$ 0.016\\
  Eccentric EB & 0.159 $\pm$ 0.019 & 0.166 $\pm$ 0.020 & 0.121 $\pm$ 0.019\\
  \hline
 \end{tabular}
\end{table}
\begin{figure}
 \includegraphics[width=\columnwidth]{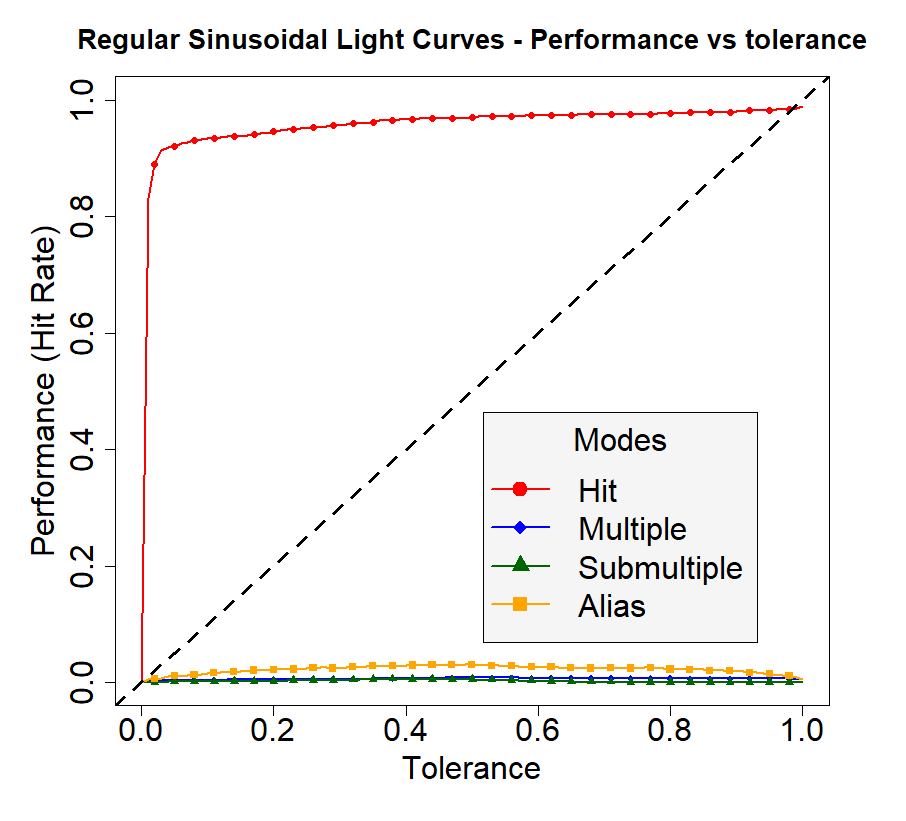}
 \caption{Performance vs Tolerance for the regular cadence sinusoidal light curves. The hit rate rapidly rises to nearly perfect very quickly showing GRAPE can effectively fine tune sinusoidal periods.}
 \label{fig:regcadsin}
\end{figure}

\begin{figure}
 \includegraphics[width=\columnwidth]{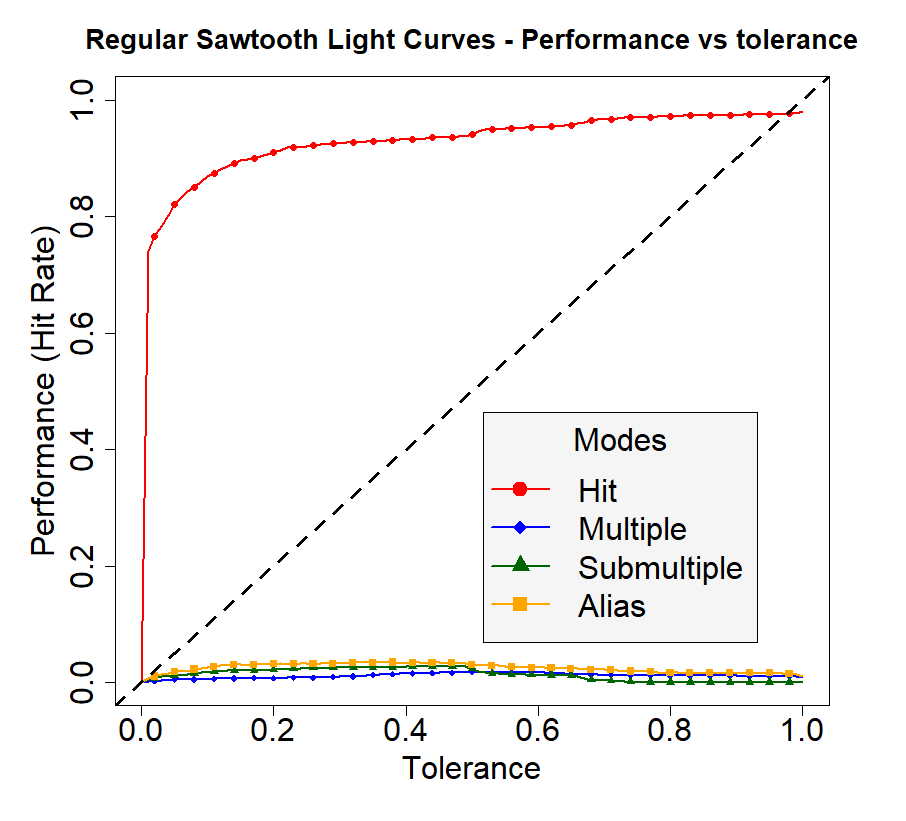}
 \caption{Performance vs Tolerance for the regular cadence sawtooth light curves. The hit rate rise is similar to the sinusoidal light curves but with an increased number of multiple and aliased periods.}
 \label{fig:regcadsaw}
\end{figure}

\begin{figure}
 \includegraphics[width=\columnwidth]{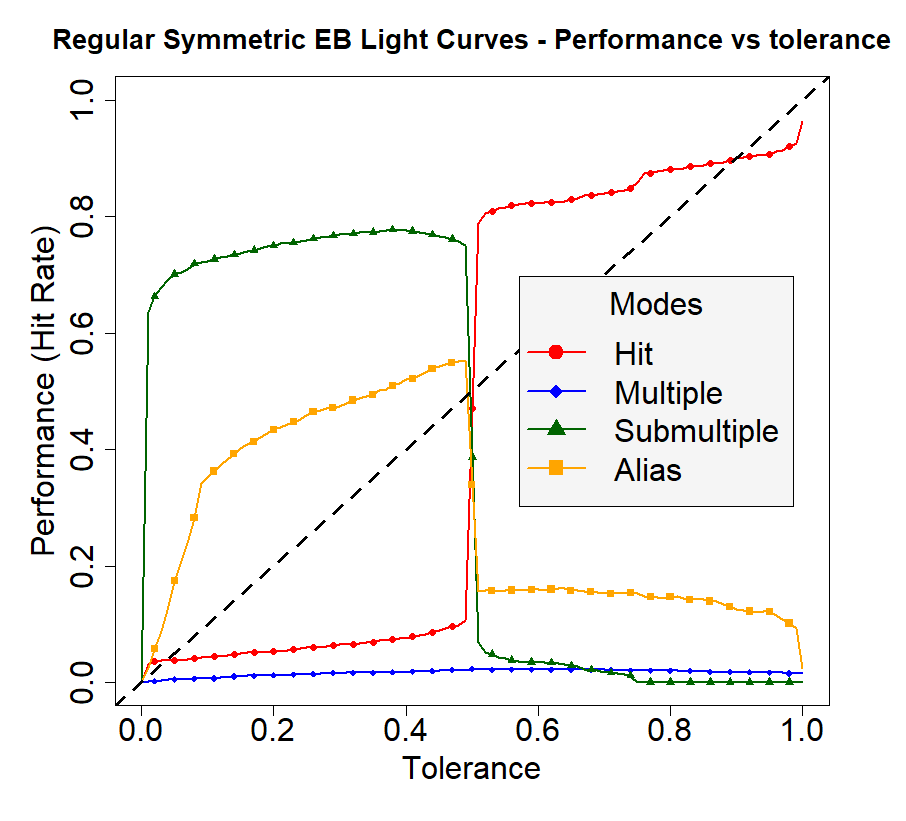}
 \caption{Performance vs Tolerance for the regular cadence symmetric eclipsing binaries. The $N = 2$ submultiple is the dominant failure mode until the tolerance reaches $\epsilon = 0.5$. Here, the submultiple satisfies the hit inequality and causes the seen performance swap. This failure mode is caused by the secondary eclipse at phase 0.5. The sinusoidal model provides a better hit at half the true period through the combined eclipses as it is unable to sufficiently model the differentially sized eclipses at the true period.}
 \label{fig:regcadecl}
\end{figure}

\begin{figure}
 \includegraphics[width=\columnwidth]{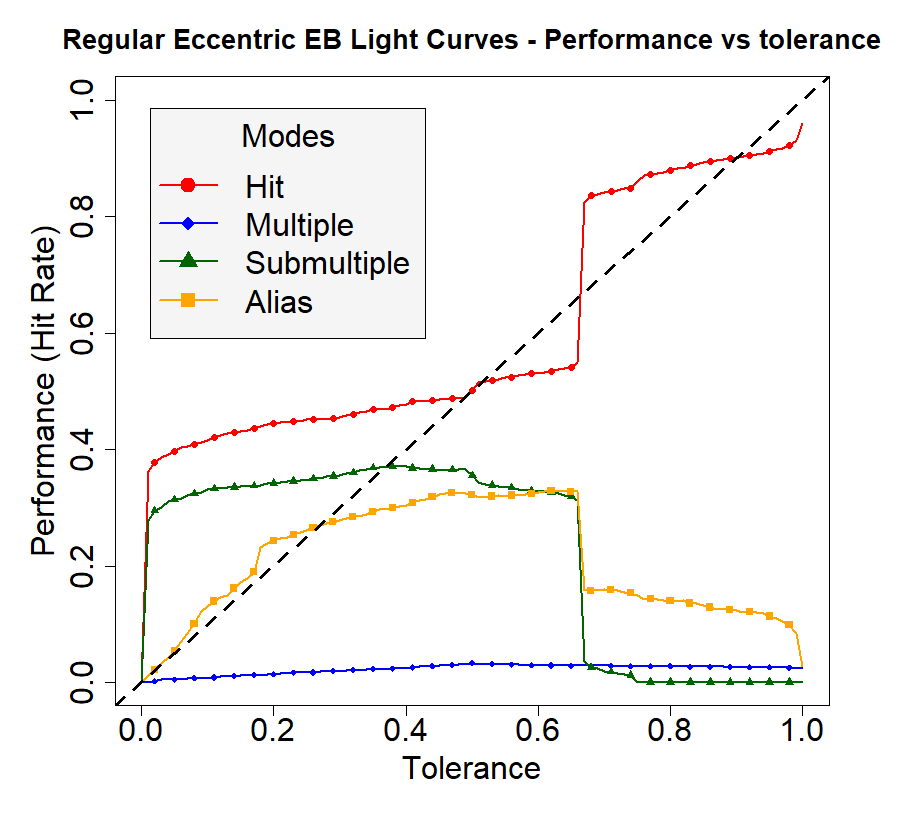}
 \caption{Performance vs Tolerance for the regular cadence eccentric eclipsing binaries. Here about half the non-aliased light curves are hits and $N = 3$ submultiples. This is a result of the eccentric secondary eclipse being at 0.7, close to 0.667. Therefore, depending on the sampling, some light curves appear as a third of the period with a missing third eclipse or else the eccentricity of the phased light curve results in the best fit being at the true period. This is only expected to happen for secondary eclipse phases near 0.333 and 0.667 as half the eclipses would be missing for the $N = 4$ submultiple variant at phases 0.25 or 0.75 which would prevent period matching at a quarter of the true period.}
 \label{fig:regcadeeb}
\end{figure}

GRAPE clearly outperforms the BGLS periodogram in the correct period estimation of the sinusoidal light curves with a hit rate improvement of 10\%. Sawtooth light curves performed better on GRAPE by 2\% for the regular cadence light curves and similarly on the Skycam cadence light curves as well as identifying some poorer quality light curves as aliases. The symmetric and eccentric eclipsing binaries suffer from a significant submultiple failure mode. This is a well understood result of the Lomb-Scargle method and its extensions. GRAPE maintains the performance of the BGLS periodogram within the statistical confidence levels for the symmetric eclipsing binaries. The eccentric eclipsing binaries were the worst performing shape on both GRAPE and the BGLS periodogram as they were the least sinusoidal of the light curves and therefore the BGLS fitness function is not tuned for them. The periodogram slightly outperformed GRAPE for this shape of light curve. This is likely due to the importance of the fitness function in the propagation of genetic information. For eccentric binaries, the response from the correct period did not outperform a poor period by a significant margin and therefore it was never optimised heavily, whereas the frequency spectrum would sample a period close to the true period by default. These failings are not a required disadvantage of GRAPE. It is entirely possible to replace the fitness function with another more appropriate to the required task and achieve the same performance increase as the sinusoidal light curves did with the BGLS fitness function. Ultimately, this experiment shows that GRAPE exhibits a similar performance to the BGLS periodogram with Vuong Closeness for every light curve shape other than purely sinusoidal. This demonstrates that the performance of the method is primarily driven by the Vuong Closeness test and its ability to distinguish between hit, multiple and aliased models. The frequency spectrum approach did not appear to result in the expected loss of performance. It is likely that the use of the multiple models in the Vuong Closeness test corrected for this weakness.

We also decided to investigate the tolerance of the two methods. For the previous experiment we produced results with a tolerance $\epsilon = 0.01$. It is desirable to achieve high hit-rates at as low a tolerance as possible but as GRAPE has been designed for use in a classification pipeline, we can still generate informative light curve features if additional light curves require a higher tolerance. Figures \ref{fig:regcadsin} to \ref{fig:regcadeeb} show plots of the tolerance of $0.0 \leqslant \epsilon \leqslant 1.0$ against the recovered rate of hits, multiples, submultiples and aliases for the four different light curve shapes with the regular cadence. The Skycam cadence light curves show a much higher incidence rate of aliases due to the poorer phase sampling of some periods.

\begin{table}
 \caption{GRAPE period estimation results with a seed of 1 on regular cadence simulated light curves with a tolerance $\epsilon = 0.01$.}
  \label{tab:grape1}
 \begin{tabular}{lccc}
  \hline
  Type & Hit & Multiple & Alias\\
  \hline
  Sinusoidal & 0.833 $\pm$ 0.019 & 0.001 $\pm$ 0.002 & 0.000 $\pm$ 0.000\\
  Sawtooth & 0.754 $\pm$ 0.023 & 0.004 $\pm$ 0.004 & 0.004 $\pm$ 0.003\\
  Symmetric EB & 0.026 $\pm$ 0.009 & 0.647 $\pm$ 0.025 & 0.025 $\pm$ 0.010\\
  Eccentric EB & 0.371 $\pm$ 0.025 & 0.279 $\pm$ 0.024 & 0.007 $\pm$ 0.005\\
  \hline
 \end{tabular}
\end{table}

\begin{table}
 \caption{GRAPE period estimation results with a seed of 2 on regular cadence simulated light curves with a tolerance $\epsilon = 0.01$.}
  \label{tab:grape2}
 \begin{tabular}{lccc}
  \hline
  Type & Hit & Multiple & Alias\\
  \hline
  Sinusoidal & 0.823 $\pm$ 0.020 & 0.001 $\pm$ 0.002 & 0.009 $\pm$ 0.006\\
  Sawtooth & 0.727 $\pm$ 0.023 & 0.009 $\pm$ 0.005 & 0.004 $\pm$ 0.004\\
  Symmetric EB & 0.037 $\pm$ 0.010 & 0.619 $\pm$ 0.025 & 0.026 $\pm$ 0.009\\
  Eccentric EB & 0.352 $\pm$ 0.025 & 0.286 $\pm$ 0.024 & 0.010 $\pm$ 0.007\\
  \hline
 \end{tabular}
\end{table}

\begin{table}
 \caption{GRAPE period estimation results with a seed of 3 on regular cadence simulated light curves with a tolerance $\epsilon = 0.01$.}
  \label{tab:grape3}
 \begin{tabular}{lccc}
  \hline
  Type & Hit & Multiple & Alias\\
  \hline
  Sinusoidal & 0.838 $\pm$ 0.019 & 0.001 $\pm$ 0.002 & 0.004 $\pm$ 0.004\\
  Sawtooth & 0.741 $\pm$ 0.023 & 0.005 $\pm$ 0.004 & 0.011 $\pm$ 0.007\\
  Symmetric EB & 0.032 $\pm$ 0.009 & 0.644 $\pm$ 0.025 & 0.021 $\pm$ 0.008\\
  Eccentric EB & 0.364 $\pm$ 0.025 & 0.274 $\pm$ 0.023 & 0.017 $\pm$ 0.009\\
  \hline
 \end{tabular}
\end{table}

Due to the random seeds used in GRAPE, we repeated the experiment with three different seed values $\mathrm{Seed}_D = [1,2,3]$ to see how the performance varied with the random processes inside a genetic algorithm. Table \ref{tab:grape1} shows the performance of the $\mathrm{Seed}_D = 1$ period estimation on the 1000 synthetic light curves with table \ref{tab:grape2} showing $\mathrm{Seed}_D = 2$ performance and table \ref{tab:grape3} showing $\mathrm{Seed}_D = 3$ performance. The reported errors are the 95\% confidence intervals as computed by a bootstrapping confidence estimator with 100,000 resamples. For the modes with sufficient population to determine an accurate confidence interval, the confidence intervals indicate that the results of GRAPE are consistent over a large set of light curves. Whilst the performance of an individual light curve can vary depending on seed, the overall percentage of matched light curves should remain consistent for a large dataset. For important individual light curves, GRAPE can be rerun with different seeds or alternatively, with a larger set of returned candidate periods as this will reduce the chance that a good trial period is not detected and evaluated.

We also computed a stratified subset of 400 of the light curves, 100 for each shape. This stratified set was generated two more times in addition to the previous set resulting in three different additive noise generations to determine the consistency of the period estimation performance. This was performed for the Skycam cadence light curves. The confidence intervals are again the 95\% confidence intervals as determined from a bootstrapping estimation using 100,000 resamples. The results are shown in table \ref{tab:noisecons} with the Data column indicating the period estimation algorithm and the additive noise seed number $[10,20,30]$ utilised in the bootstrap. As with the GRAPE seed performance, the different noise models are consistent to the 95\% confidence intervals. Therefore, we suggest that the performance of GRAPE and the BGLS periodogram are consistent for large datasets independent of the random seeds used for either noise generation or genetic algorithm operation.
\begin{table}
 \caption{GRAPE and BGLS periodogram period estimation results with three different light curve additive noise models on Skycam cadence simulated light curves with a tolerance $\epsilon = 0.01$.}
  \label{tab:noisecons}
 \begin{tabular}{lccc}
  \hline
  Data & Hit & Multiple & Alias\\
  \hline
  GRAPE lcseed 10 & 0.348 $\pm$ 0.040 & 0.128 $\pm$ 0.028 & 0.073 $\pm$ 0.023\\
  GRAPE lcseed 20 & 0.363 $\pm$ 0.040 & 0.120 $\pm$ 0.028 & 0.098 $\pm$ 0.023\\
  GRAPE lcseed 30 & 0.365 $\pm$ 0.040 & 0.103 $\pm$ 0.025 & 0.080 $\pm$ 0.023\\
  BGLS lcseed 10 & 0.310 $\pm$ 0.038 & 0.128 $\pm$ 0.028 & 0.090 $\pm$ 0.028\\
  BGLS lcseed 20 & 0.313 $\pm$ 0.038 & 0.110 $\pm$ 0.025 & 0.090 $\pm$ 0.025\\
  BGLS lcseed 30 & 0.335 $\pm$ 0.040 & 0.103 $\pm$ 0.025 & 0.095 $\pm$ 0.030\\
  \hline
 \end{tabular}
\end{table}

\begin{figure}
 \includegraphics[width=\columnwidth]{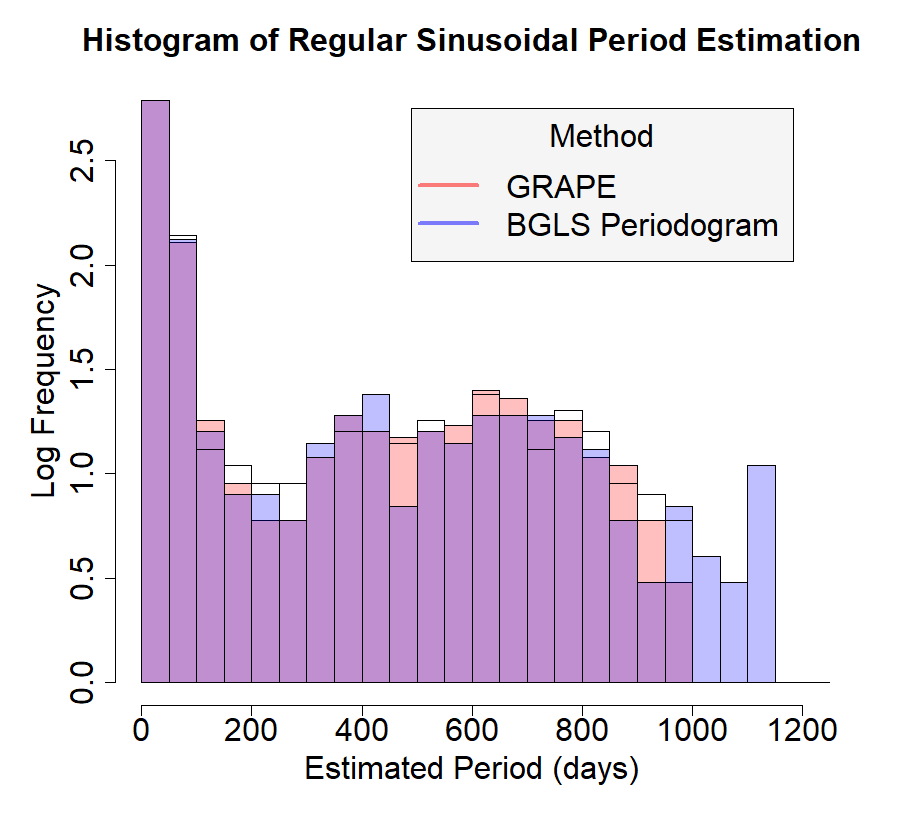}
 \caption{Histogram of the base-10 logarithm of the frequency of light curves with a given period as a function of the estimated period for the regular cadence sinusoidal light curves. The white regions indiciate that there were more initial light curves in this period range than identified by either algorithm, it is the initial configuration of the test periods. For the sinusoidal light curves, performance was good from both fitness functions, as seen in figure 6. GRAPE performs better at longer periods due to the frequency spectrum sampling this region poorly. The overabundance of BGLS periodogram results around 900-1200\,days are likely Vuong-reflected submultiples of the poorly detected longer periods.}
 \label{fig:histsin}
\end{figure}

\begin{figure}
 \includegraphics[width=\columnwidth]{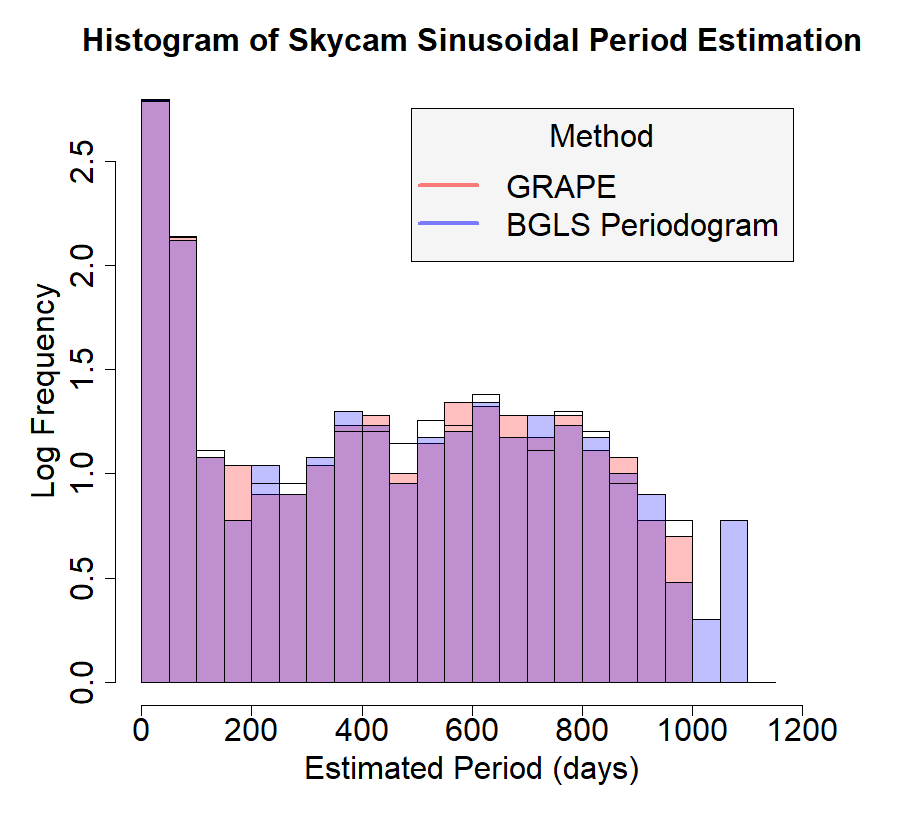}
 \caption{Histogram of the base-10 logarithm of the frequency of light curves with a given period as a function of the estimated period for the Skycam cadence sinusoidal light curves. The performance on the Skycam cadence sinusoids is similar to that of the regular cadence sinusoids. This is an interesting result as it indicates that the uneven sampling of Skycam may not be significantly detrimental to continuous variations.}
 \label{fig:histsinskt}
\end{figure}

\begin{figure}
 \includegraphics[width=\columnwidth]{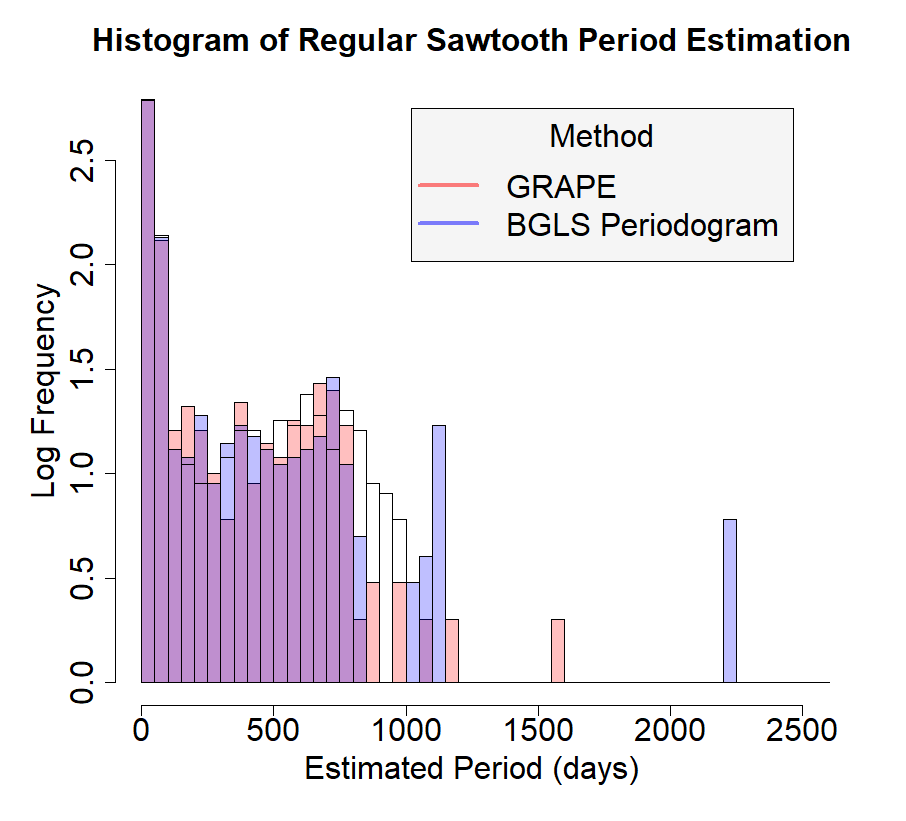}
 \caption{Histogram of the base-10 logarithm of the frequency of light curves with a given period as a function of the estimated period for the regular cadence sawtooth light curves. Many very long estimated periods are obvious due to mispredictions by the Vuong closeness test in both GRAPE and the BGLS periodogram. There are also many spurious detections near the time span of the light curves for this test despite being on regular cadence as the sawtooth shape diverges from the expectations of the BGLS fitness function.}
 \label{fig:histsaw}
\end{figure}

\begin{figure}
 \includegraphics[width=\columnwidth]{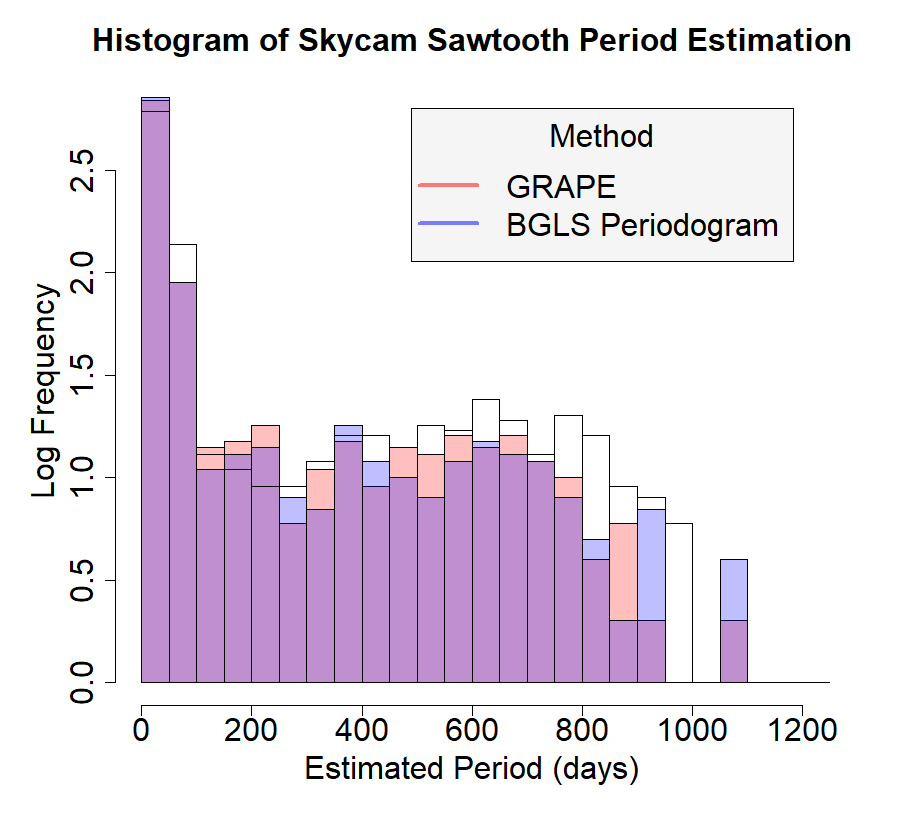}
 \caption{Histogram of the base-10 logarithm of the frequency of light curves with a given period as a function of the estimated period for the Skycam cadence sawtooth light curves. The performance of the Skycam cadence sawtooth light curves appears similar to the regular cadence. There is a noticible lack of period estimations near 1000\,days likely due to sampling difficulties with a single sawtooth variation.}
 \label{fig:histsawskt}
\end{figure}

Our next experiment is to determine if the failure modes of GRAPE and the BGLS periodogram have a dependence on the period. In the frequency spectrum case of the BGLS periodogram, we would expect to see the performance of the periodogram be a function of the candidate period. We therefore plot two histograms of the GRAPE estimated periods and the periodogram estimated periods on the sinusoidal light curves and is displayed in figure \ref{fig:histsin}. The histogram clearly shows that both methods perform more poorly the closer the period is to $P_{max}$. This is due to poorer sampling of the phase space as there are less complete cycles viewed inside of the light curve baseline. Additionally, the BGLS periodogram suffers a much greater performance loss at this extreme. This is likely due to a combination of the frequency spectrum selecting a submultiple of the true period followed by a Vuong Closeness test correction. As this correction must be an integer multiple of the initially detected period, this introduces an error near $P_{max}$. This can be seen in figure \ref{fig:histsin} as a selection of light curves with BGLS periodogram estimated periods between 1000 and 1200 days. GRAPE performs much better in this range due to treating the parameter space as a continuous variable whereas the periodogram samples the high periods extremely sparsely and thus GRAPE does not need to rely on the Vuong Closeness test to correct as many long periods. Figure \ref{fig:histsinskt} demonstrates the same histogram for the Skycam cadence sinusoidal light curves. The performance on the Skycam cadence sinusoidal light curves was similar to the regular cadence sinusoidal light curves except with a larger instance of aliased periods due to poor sampling. This is an interesting result and demonstrates that the Skycam sampling does not adversely effect the performance of the period estimation when the fitness function is a good match to the data. Figure \ref{fig:histsaw} shows the same histogram for the regular cadence sawtooth light curves and figure \ref{fig:histsawskt} shows the histogram of the Skycam cadence sawtooth light curves. The same effects can be seen but with additional extremely long periods found by GRAPE and the BGLS periodogram due to the vuong closeness test deciding to pick a multiple of the identified period for many light curves with periods between 850-1000\,days. This effect appears worse in the regular cadence light curves with the Skycam cadence light curves showing a higher rate of aliases instead. This is likely due to the tested model being a simple sinusoid which does not fit the sawtooth signal combined with a minimal number of observed cycles. A more generalised model for the Vuong closeness test would be desirable for this event especially if a different non-sinusoidal fitness function is selected.

\begin{figure}
 \includegraphics[width=\columnwidth]{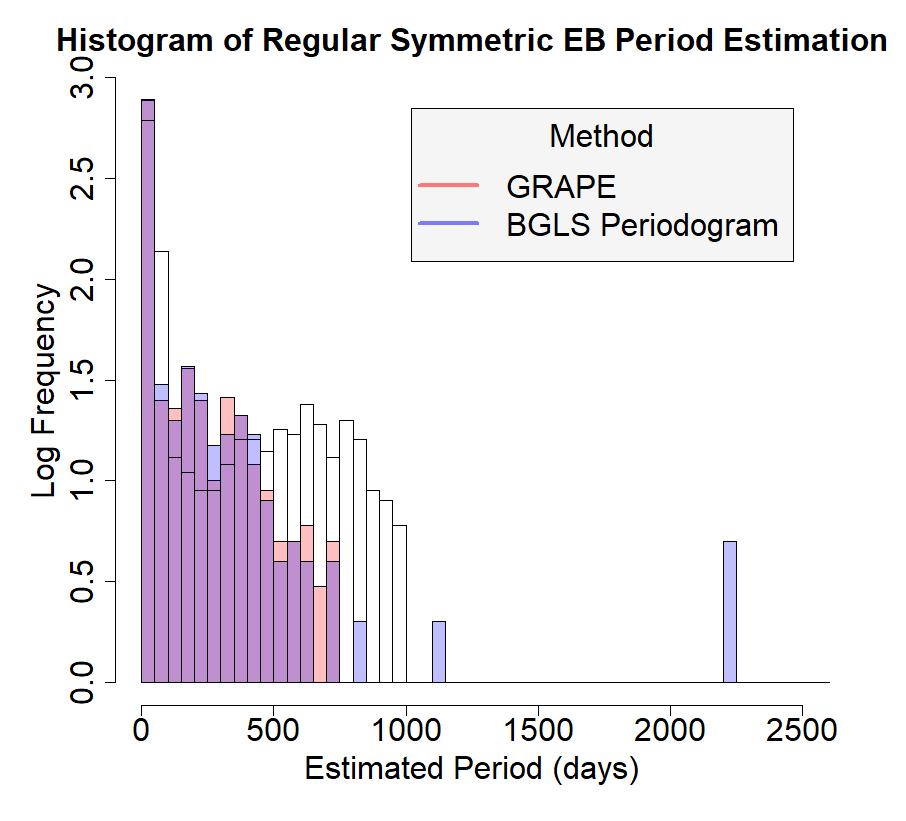}
 \caption{Histogram of the base-10 logarithm of the frequency of light curves with a given period as a function of the estimated period for the regular cadence symmetric eclipsing binaries. There is a significant depletion in the long periods after 500\,days due to the common $N = 2$ failure mode. Some of the longer periods are correctly estimated possibly due to sampling resulting in no data points from the secondary eclipse. In this case, the best fitting sinusoidal model is at the correct period.}
 \label{fig:histecl}
\end{figure}

\begin{figure}
 \includegraphics[width=\columnwidth]{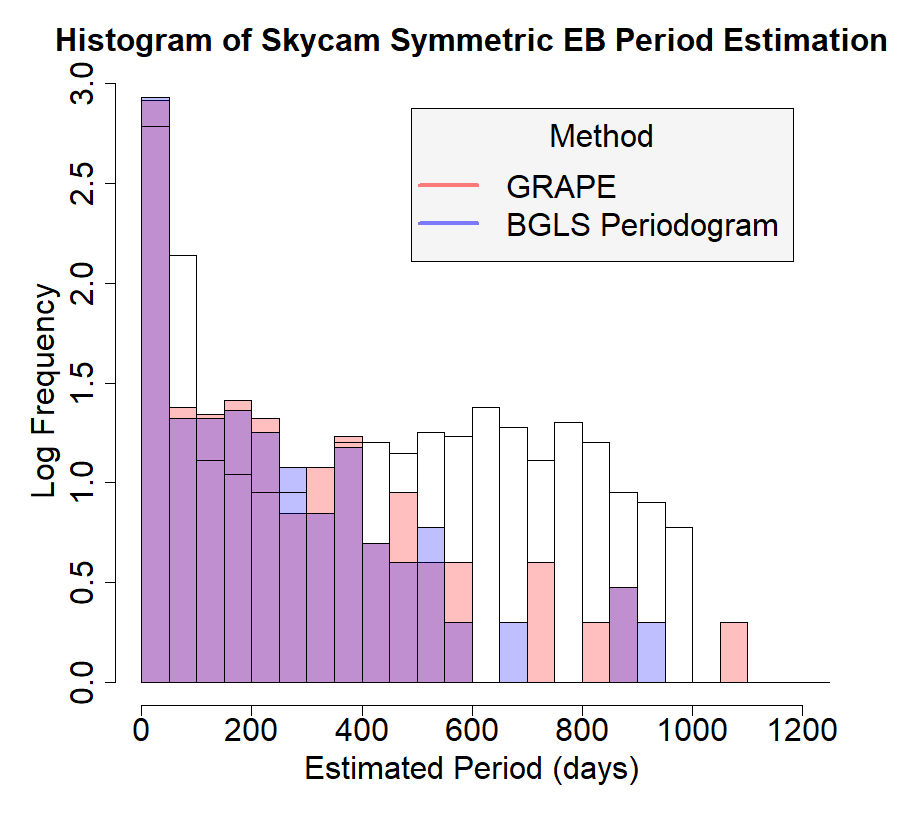}
 \caption{Histogram of the base-10 logarithm of the frequency of light curves with a given period as a function of the estimated period for the Skycam cadence symmetric eclipsing binaries. The Skycam cadence light curves have similar performance to the regular cadence light curves except for additional long period hits. This is likely due to the even greater chance of loosing either eclipse at long periods due to the sampling rate of Skycam cadence compared to the regular cadence.}
 \label{fig:histeclskt}
\end{figure}

\begin{figure}
 \includegraphics[width=\columnwidth]{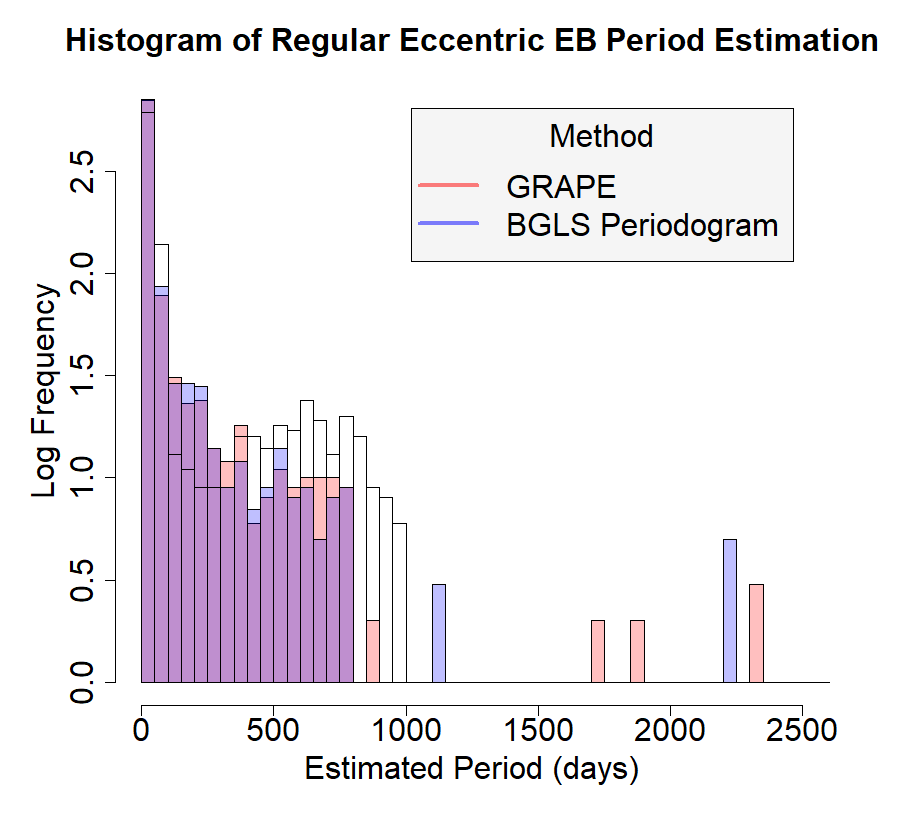}
 \caption{Histogram of the base-10 logarithm of the frequency of light curves with a given period as a function of the estimated period for the regular cadence eccentric eclipsing binaries. Many of the light curves have been underestimated into the $N=3$ submultiple. This results in an overabundance of periods below 500\,days. About half of the period estimates are correct and the other half are in this $N=3$ submulitple failure mode.}
 \label{fig:histeeb}
\end{figure}

\begin{figure}
 \includegraphics[width=\columnwidth]{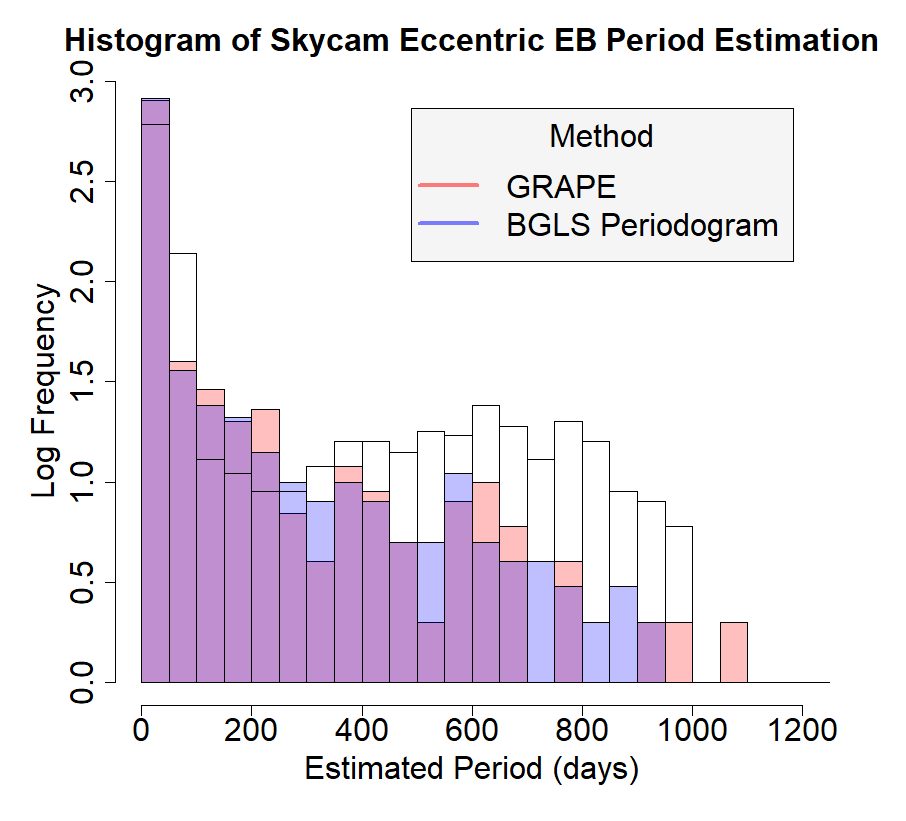}
 \caption{Histogram of the base-10 logarithm of the frequency of light curves with a given period as a function of the estimated period for the Skycam cadence eccentric eclipsing binaries. The poor performance at detecting many of the long period light curves is due to limits in the BGLS fitness function at fitting eccentric eclipsing binaries. Additionally, the Skycam cadence results in many of the eclipses being unobserved resulting in a spurious period estimation.}
 \label{fig:histeebskt}
\end{figure}

The symmetric eclipsing binary histograms in figures \ref{fig:histecl} and \ref{fig:histeclskt} show similar errors to the sawtooth light curves with poor Vuong Closeness estimating periods outside of the long period range. The periods are highly underestimated above 500 days due to the common $N = 2$ submultiple failure mode of symmetric eclipsing binaries. The eccentric eclipsing binaries are split almost equally between hits and the $N = 3$ submultiple showing a moderate long period depletion. The eccentric eclipsing binaries perform extremely poorly at long periods in both regular and Skycam cadence as seen in figures \ref{fig:histeeb} and \ref{fig:histeebskt}. This is a result of many of the long periods exhibit the $N = 3$ failure mode due to the small number of sampled eclipses. Additionally, at long periods the phase sampling often fails to measure the eclipse at all due to the seasonal sampling windows we have added to our light curves, especially for the Skycam cadence. This is a common issue in real survey eclipsing binaries \citep{b48,b49} and arguably a bigger problem as our simulated light curves have unphysically large eclipse durations at long periods which decrease the probability that the eclipse will be missed. It is interesting to note that the Vuong Closeness test long period multiplication failure mode does not seem to occur on the Skycam cadence light curves. This is possibly due to the poor long duration sampling producing poor quality models that do not improve over the initial period estimation model. Alternatively, the alias of the period might be selected instead due to the strong Skycam aliases.

\subsection{Cadence-dependent performance}
\begin{figure}
 \includegraphics[width=\columnwidth]{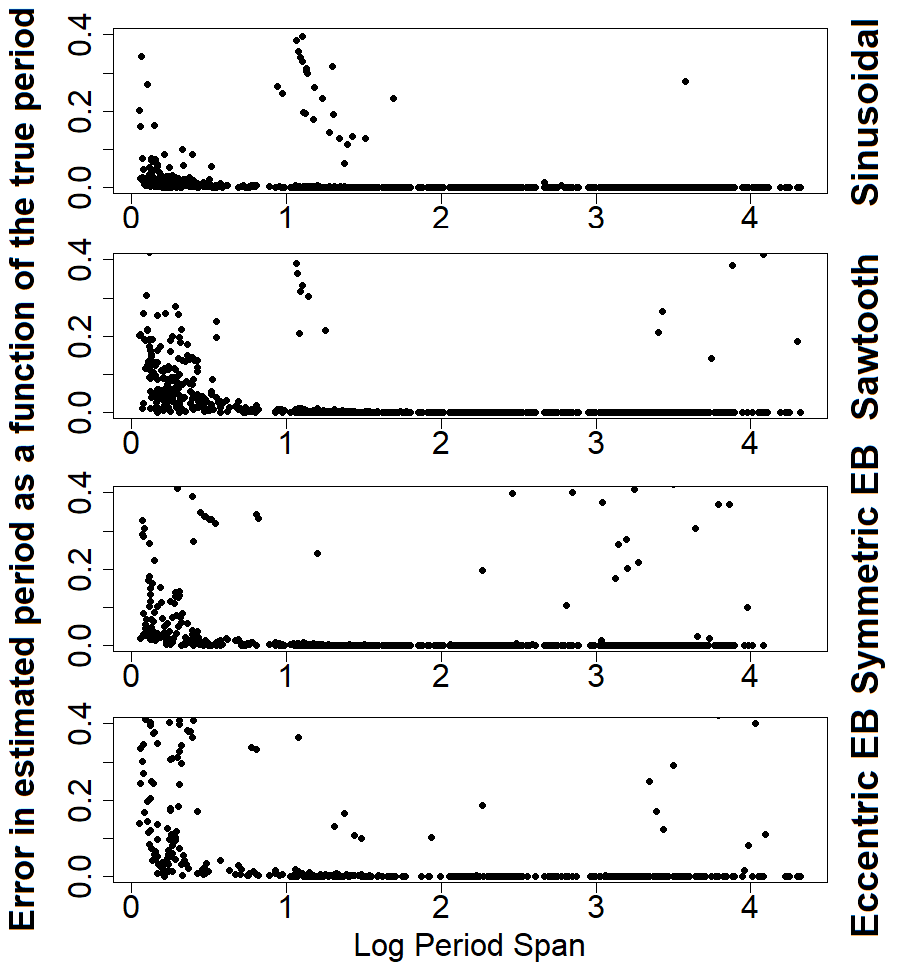}
 \caption{Plot of the base-10 logarithm of the period span of light curves with a given period as a function of the estimated period fractional error for the regular cadence light curves.}
 \label{fig:perspanreg}
\end{figure}

\begin{figure}
 \includegraphics[width=\columnwidth]{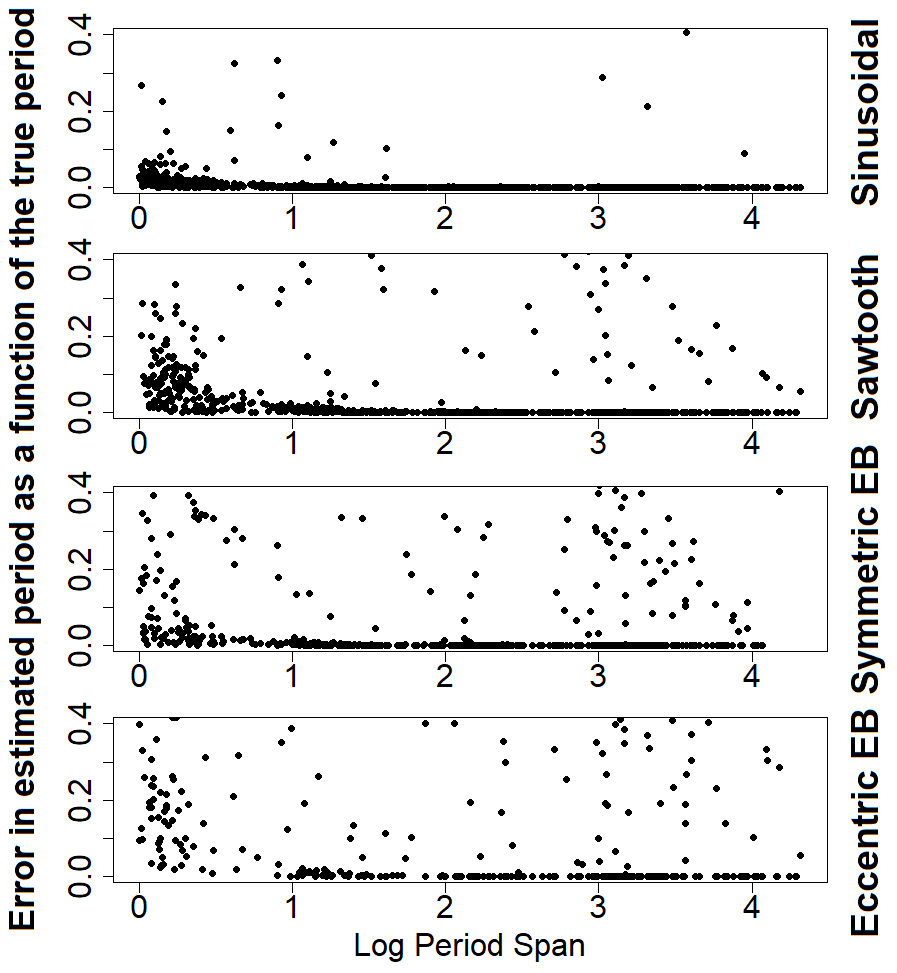}
 \caption{Plot of the base-10 logarithm of the period span of light curves with a given period as a function of the estimated period fractional error for the Skycam cadence light curves.}
 \label{fig:perspanskt}
\end{figure}

The error in the estimated periods against the period span is an excellent indicator of the performance of the GRAPE period estimation due to cadence against the simulated period. Period span is defined as the baseline of the light curves divided by the input period and is the number of cycles present in the light curve. It is calculated by Equation \ref{eq:pspan}.
\begin{equation}
P_{span} = \frac{t_{max} - t_{min}}{P_{i}}
\label{eq:pspan}
\end{equation}
where $t_{min}$ is the minimum time instant of the light curve, $t_{max}$ is the maximum time instant of the light curve and $P_{i}$ is the input period. For low values of $P_{span}$, the performance is expected to be poorer as there are less complete cycles and the cadence results in unsampled regions of the phase space. Large performance errors occuring where $P_{span} = t_{max} - t_{min}$ also indicate that the cadence is resulting in substantial aliasing. The estimated error is calculated by Equation \ref{eq:esterr} and is the fractional error from the input period.
\begin{equation}
\xi = \frac{|P_{i} - P_{e}|}{P_{i}}
\label{eq:esterr}
\end{equation}
Figure \ref{fig:perspanreg} shows this plot for the regular cadence light curves of the four types and figure \ref{fig:perspanskt} shows the results of GRAPE on the Skycam cadence light curves. Symmetric eclipsing binaries estimated periods have been doubled due to the common Lomb-Scargle failure mode. The regular cadence light curves show only two main patterns. The increase in error near $\log_{10}\left(P_{\mathrm{span}}\right) = 0$ is due to the poorer sampling of the shape of the light curve. This is less of an issue in the sinusoidal light curves as the model can interpolate the missing data but it becomes progressively a bigger problem as the light curve shape becomes more non-sinusoidal. Missing the eclipses in the eclipsing binary light curves also causes smaller $P_{span}$ values to perform poorly. There is also a number of failed period estimations near $\log_{10}\left(P_{\mathrm{span}}\right) = 1$ due to the seasonal sampling periodicity as the Vuong Closeness test incorrectly determines multiples of the period due to unsampled data.

The Skycam cadence plot in Figure \ref{fig:perspanskt} exhibits the same patterns but with an additional uncorrelated set of erroneous period estimations due to the sampling of these light curves being insufficient despite many cycles being present in the baseline. The dominant feature of this failure mode is the increased period estimation errors near $\log_{10}\left(P_{\mathrm{span}}\right) = 3$ which is close to the sidereal day spurious period. The light curves of this additional failure mode also increases in number as the underlying signal becomes more non-sinusoidal. It is clear that the Skycam sampling appears inferior to the regular cadence yet this analysis can be used to augment the spurious and alias correction steps present in GRAPE. Ultimately, the advantage in Skycams capability of performing survey astronomy independently of the operations of the Liverpool Telescope is met with the significant disadvantage that many variable objects observed by the instruments will not be sampled at sufficient quality to detect.

\subsection{Runtime considerations}

\begin{figure}
 \includegraphics[width=\columnwidth]{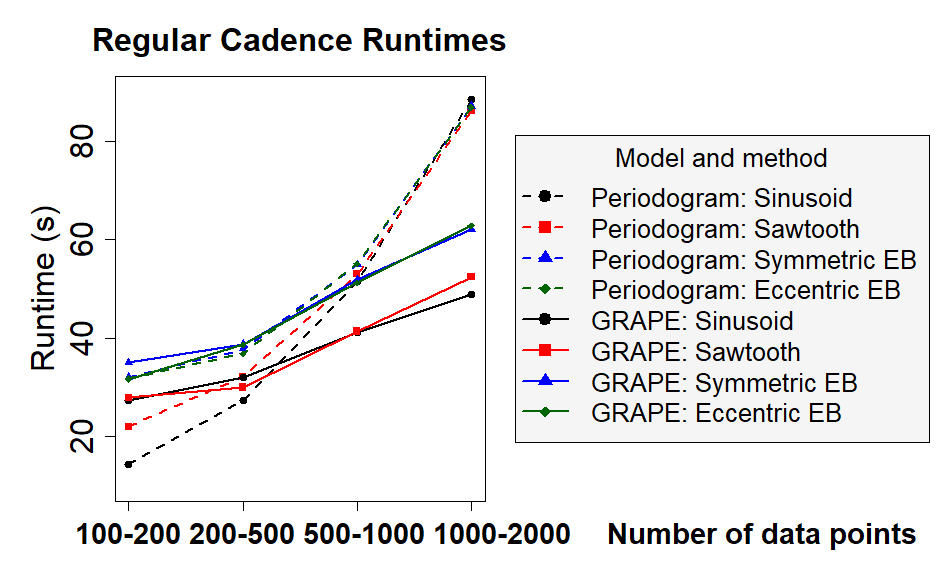}
 \caption{Plot of the average runtime of the different shaped light curves as a function of the number of data points for the regular cadence light curves.}
 \label{fig:regcadruntimes}
\end{figure}

The previous experiments have shown the GRAPE performance to be consistent with the BGLS periodogram and even improved for signals close to the modelled fitness function. This fulfills the first major requirement for this method as an application to the Skycam survey data. The second requirement, which is also the original driving force behind the development of GRAPE, is the requirement for the period estimation task to be as computationally efficient as possible. Many of the important properties of the Skycam light curves are extracted from statistics which are a function of the candidate period and therefore require the estimation of a period prior to calculation. For large numbers of light curves this calculation must be as rapid as possible whilst maintaining performance. To understand the runtime requirements of GRAPE compared to a periodogram approach we calculated the average light curve processing time by calculating the mean runtime for the stratified set of 400 regular cadence light curves and 400 Skycam cadence light curves used in the testing of the additive noise models. These light curves are seperated by the number of data points they contain based on the intervals selected when generating the synthetic data. Figure \ref{fig:regcadruntimes} shows the mean runtime in seconds of the groups of light curves against the binned number of data points for the regular cadence dataset seperated by light curve shape and period estimation method. A number of interesting patterns emerges in this result. Firstly, the periodogram has an exponential dependency on the number of data points compared to GRAPE. This is expected as the periodogram algorithm is an $O(N^2)$ method whereas GRAPE, whilst having an additional overhead due to the generational updates which are not a function of data point number, is an $O(N)$ method. As both methods use the Vuong Closeness test, its contribution is present in both operations. The effect of Vuong Closeness is visible in the form of the runtime separation of the different light curve shapes. For the periodogram, this results in sinusoidal and sawtooth light curves requiring less runtime at low numbers of data points compared with eclipsing binary light curves. At higher numbers of data points, the runtime required by the $O(N^2)$ periodogram becomes dominant and all the runtimes converge. For GRAPE, the effect is to also separate the sinusoidal and sawtooth runtimes from the slower eclipsing binary runtimes. However, due to the linear dependence on data points of GRAPE, this runtime difference remains to higher data point light curves.

\begin{figure}
 \includegraphics[width=\columnwidth]{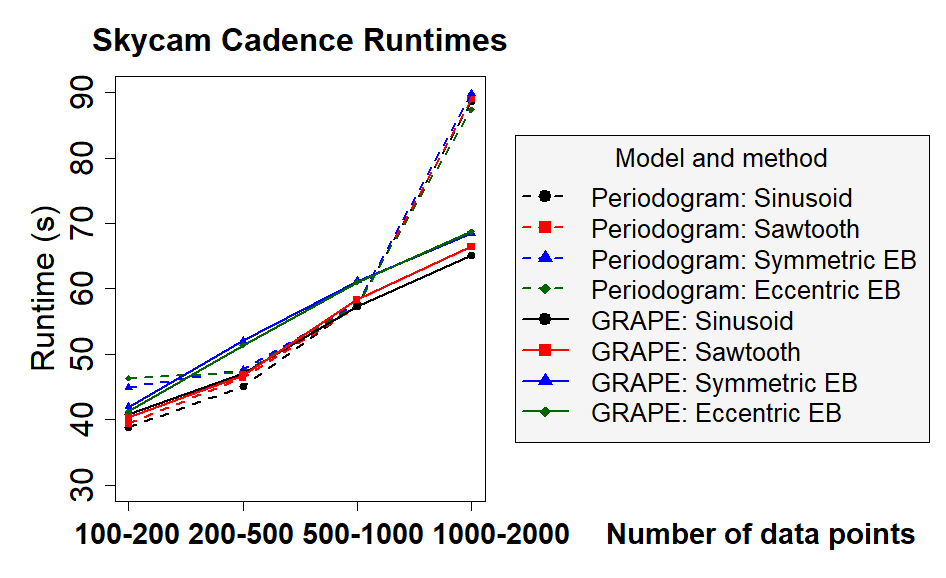}
 \caption{Plot of the average runtime of the different shaped light curves as a function of the number of data points for the Skycam cadence light curves.}
 \label{fig:sktcadruntimes}
\end{figure}

For the Skycam cadence results shown in figure \ref{fig:sktcadruntimes}, the runtimes are longer and similar effects to the regular cadence light curves are seen but are suppressed. This suppression is likely a result of the poorer sampling quality of the light curves. This results in increased computational effort required by both GRAPE genetic optimisation and the Vuong Closeness test used in both GRAPE and the periodogram. As a result, the periodogram requires more time to process the lower data point light curves as the aliased periods are more dominant. The separation of the runtimes on the sinusoidal and sawtooth light curves compared to the eclipsing binary light curves is still apparent but reduced as the sinusoidal and sawtooth light curves require additional computation to deal with aliased periods. Ultimately, for the Skycam cadence light curves, which is of more interest due to them more closely reproducing the properties of the real survey data, GRAPE appears to maintain similar performance to the periodogram for the sinusoidal and sawtooth light curves and requires less runtime than the periodogram on light curves with greater than $500-1000$ data points. Whilst many of the interesting Skycam light curves do have data points in this range, there are many light curves with data points from $1000-15,000$ data points which would require an unacceptably long time to run using a periodogram (due to the $O(N^2)$ requirement of this method). GRAPE provides an approach to obtain the desired results in less time.


\begin{table}
 \caption{GRAPE and BGLS periodogram period estimation results on 100 stratified regular cadence sinusoidal light curves with various fine-tuning oversampling runs. All non-hit estimated periods are of unknown relation.}
  \label{tab:oversamp}
 \begin{tabular}{lccc}
  \hline
  Data & Hit & Multiple & Alias\\
  \hline
  GRAPE & 0.803 $\pm$ 0.037 & 0.000 $\pm$ 0.000 & 0.000 $\pm$ 0.000\\
  BGLS Periodogram & 0.650 $\pm$ 0.080 & 0.000 $\pm$ 0.000 & 0.000 $\pm$ 0.000\\
  $\mathrm{ofac}=20$ finetune & 0.810 $\pm$ 0.070 & 0.000 $\pm$ 0.000 & 0.000 $\pm$ 0.000\\
  $\mathrm{ofac}=50$ finetune & 0.820 $\pm$ 0.070 & 0.000 $\pm$ 0.000 & 0.000 $\pm$ 0.000\\
  \hline
 \end{tabular}
\end{table}
\begin{table}
 \caption{GRAPE and BGLS periodogram period estimation results on 100 stratified Skycam cadence sinusoidal light curves with various fine-tuning oversampling runs. There are no multiple estimated periods in this dataset.}
  \label{tab:oversampskt}
 \begin{tabular}{lccc}
  \hline
  Data & Hit & Multiple & Alias\\
  \hline
  GRAPE & 0.780 $\pm$ 0.040 & 0.000 $\pm$ 0.000 & 0.013 $\pm$ 0.013\\
  BGLS Periodogram & 0.660 $\pm$ 0.080 & 0.000 $\pm$ 0.000 & 0.050 $\pm$ 0.040\\
  $\mathrm{ofac}=20$ finetune & 0.760 $\pm$ 0.070 & 0.000 $\pm$ 0.000 & 0.020 $\pm$ 0.030\\
  $\mathrm{ofac}=50$ finetune & 0.790 $\pm$ 0.070 & 0.000 $\pm$ 0.000 & 0.010 $\pm$ 0.020\\
  \hline
 \end{tabular}
\end{table}

We also peformed one final experiment to determine the runtime required for the BGLS periodogram, with Vuong Closeness, to obtain similar results to GRAPE on the regular cadence and Skycam cadence sinusoidal light curves through the use of a fine-tuning step on the initial $N_t = 5$ candidate periods with a $\pm 10\%$ search radius using a boosted oversampling factor. As the $\mathrm{ofac}=5$ of the previous experiments was sufficient for all non-sinusoidal light curves, we perform this only on the set of 100 stratified regular cadence sinusoidal light curves and 100 stratified skycam cadence sinusoidal light curves. Table \ref{tab:oversamp} demonstrates the results of this experiment on the regular cadence light curves using the results of the GRAPE performance contrasted with the base periodogram performance with $\mathrm{ofac}=5$ and with two fine-tuning variants, one with $\mathrm{ofac}=20$ and one with $\mathrm{ofac}=50$. Table \ref{tab:oversampskt} demonstrates the results of this experiment on the Skycam cadence light curves with the same oversample fine-tuning. The results indicate that the periodogram can replicate the performance of GRAPE on sinusoidal light curves through an additional oversampling finetuning operation. For the regular cadence light curves this is possible with a fine-tuning step with an oversampling factor of 20. For the Skycam cadence light curves the fine-tuning oversampling factor was required to be 50. These performance gains do come at a computational cost which further effects the runtime of the periodogram compared to GRAPE. The mean light curve runtime for GRAPE on the regular cadence sample was 37.32\,seconds and the BGLS periodogram with no fine-tuning mean runtime was 45.43\,seconds. For Skycam cadence these runtimes were increased due to the higher difficulty in identifying candidate periods with a mean runtime of 52.59\,seconds on GRAPE and 57.55\,seconds on the periodogram. The fine-tuning operation with $\mathrm{ofac}=20$ increase this runtime to 54.94\,seconds on the regular cadence data and 64.25\,seconds on the Skycam cadence data. For the $\mathrm{ofac}=50$ fine-tuning operation, the computational expense increases to 56.77\,seconds on the regular cadence data and 64.46\,seconds on the Skycam cadence data. Ultimately, whilst the fine-tuning operations allow the periodogram to match the performance of GRAPE, it is a the cost of increasing the required runtime. As the fine-tuning operation is also frequency spectrum based, this additional computational effort is also $O(N^2)$ complexity meaning for light curves with many observations, the processing time will be prohibitive for real-time analysis.

\section{Discussion and Conclusion}

In this paper we introduced GRAPE, a period estimation statistic embedded in a genetic algorithm with a Vuong closeness test based alias and multiple model discrimination procedure. BGLS was selected as the period estimation statistic to be used as the fitness function. We note that other methods can be used instead as the role of the fitness function is highly modular and different measures can be combined. In our future work we intend to test other fitness functions as well as develop a machine learning method which can linearly combine multiple methods for improved performance \citep{b53}. Our experiments in this paper show that Lomb-Scargle type methods function on poorly sampled data due to sinusoidal interpolation. This does also mean that non-sinusoidal signal performance degrades rapidly. The simple sinusoidal models used as part of the Vuong closeness test method are also limited due to the presumption of a sinusoidal signal which results in failures for this test most notably for the non-sinusoidal regular cadence light curves shown in figures \ref{fig:histsaw}, \ref{fig:histecl} and \ref{fig:histeeb}. An alternative model for this period correction would be ideal and is something we will consider for future research direction.

It is also important to caution the use of Bayesian periodogram methods in the search for periodicity of generic signal shapes. Bayesian periodograms output probablistic statements based on the assumption that the data has been drawn from a sinusoidal model \citep{b33}. This results in the suppression of features in the periodogram which would normally convey information on the nature of the underlying periodicity. However, GRAPE has been designed for \textit{automated} functionality and therefore it is important that potential failure modes be tightly controlled. As a major risk of any optimisation method is becoming trapped in local minima, the suppression of aliases is highly useful in propagating the genetic information. Therefore, BGLS has been deployed in this work despite the output being unreliable in the regime of non-sinusoidal periodicity. It is possible that the use of a quantum evolutionary algorithm would afford additional protection from populations becoming trapped at insignificant local minima and therefore sever our current dependency on the Bayesian periodogram \citep{b50}.

GRAPE outperformed the periodogram frequency grid in all datasets with sinusoidal signals which are well described by the fitness function. We suggest, as discussed at the beginning, that GRAPE treating the period space as a continuous variable leads to this success as the genetic algorithm could fine tune the result. This is further supported by the fine-tuning periodogram method which used a frequency spectrum with $\mathrm{ofac} = 50$ to oversample the 10\% period range around the $N_t$ candidate periods. The sinusoidal light curves had a relative hit rate improvement of 18.2\% using GRAPE compared to the periodogram for the regular cadence data and 14.9\% for the Skycam cadence data with both methods utilising the same BGLS fitness function. This was determined by assuming that the failed matches in GRAPE were also failures in the periodogram and calculating the percentage of additional failures in the periodogram. Using the same method, we determine that the sawtooth light curves had a 6.4\% improvement on regular cadence and 3.7\% on Skycam cadence when using GRAPE although this is close to the 95\% confidence interval. The symmetric EB and eccentric EB light curves are too much of a departure from the sinusoidal assumptions of the Lomb-Scargle method and exhibited a GRAPE relative performance similar, possibly slightly inferior, to the periodogram when comparing the hit and submultiple rate. This is likely a result of the reliance of GRAPE on the genetic propagation of useful information about the period space during the evolution of the candidates. On a sinusoidal light curve, the genetic algorithm places additional candidates near the sinusoidal signal period due to the prevalence of superior fitting models in this region of the period space. The algorithm can then fine tune the resulting period from this region. For the eccentric EBs, the fitness function returns a substantially weaker response to candidate periods near the true underlying simulated period. As a result, it only requires the presence of a similar-strength false model (such as on a spurious or aliased signal) to `kick' candidate periods out of this region of the period space and removing it from the fine-tuning operation. In this case, the brute force approach of the periodogram frequency grid outperformed GRAPE purely because a candidate period close to the true period would always be sampled. These results are an overall measure of the relative performance and, as can be seen in the histograms in figures \ref{fig:histsin} to \ref{fig:histeeb}, the actual performance of the methods is strongly dependent on both the shape of the underlying light curve, the value of the true period and the baseline (total measured time) of the light curve.

Our experiments show the sampling inherent to the Skycam mode of surveying will lead to objects insufficiently sampled for successful identification however the yield looks reasonable based on the relative performance degradation between the regular cadence and Skycam cadence data for sinusoidal and sawtooth light curves. Unfortunately, the loss of eclipsing binaries will be substantial for the Skycam survey regardless of period estimation method from a sampling viewpoint. The simulated data we present in this research contains only a signal and white noise component. In reality, red (correlated) noise sources are common in real light curves. GRAPE currently makes no attempt to address the presence of red noise within candidate light curves, with the BGLS fitness function assuming purely white noise residuals. Whilst this is something that can be addressed in future work, at the moment we make use of a prewhitening technique to eliminate large correlated systematic signals when applying this method to real light curves from the Skycam database.

GRAPE has a notable computational time benefit over the frequency spectrum on light curves with more than 500-1000 data points, occasionally with increased performance likely based on the chosen fitness function. This is due to the genetic algoritms $O(N)$ dependency on the number of data points in a light curve compared with the $O(N^2)$ of the frequency spectrum approach. We found that the topology of our genetic algorithm, as defined by the arguments listed in the previous section, are close to the fastest implementation we could produce before a substantial loss in performance due to having insufficient population or generations to explore the period space. We found that the linear decay of mutation was an incredibly powerful way of maintaining performance whilst decreasing the number of generations. The mutation can be described with a thermodynamic analogy. At the beginning the mutation is extremely high and the candidate periods jump rapidly across the parameter space like hot particles escaping a nearby potential. As the mutation rate decreases, the population has less energy to climb out of the potential wells and therefore begin to fine tune the local periods compared to exploring new regions of the parameter space. Eventually the mutation rate is so low that very little to no new exploration is occuring as the population cannot stray far from the local potential. Individuals located in poor areas of the parameter space die off leaving the group near the true period to reproduce solely with the purpose of fine tuning this result. We are satisfied with the computational overhead required as many Skycam light curves do have a large number of data points which become prohibitively difficult to compute a periodogram on combined with the improved accuracy of GRAPE over the periodogram. For a Skycam light curve with 5114 data points it takes the frequency spectrum periodogram about 454.7\,seconds to complete whereas GRAPE completes the same task in 93.5\,seconds. We are working on upgrading the classical genetic algorithm used in this code to a quantum genetic algorithm \citep{b50}. This method makes use of a quantum population which encodes the candidate periods as a probability density with a centroid and width. This allows for a smaller population to accomplish the same exploration and fine tuning of the parameter space decreasing the runtime further.

GRAPE is currently specifically designed to identify purely periodic phenomena. There are many quasi-periodic and semiregular variable objects in astronomy. These objects suffer a significance degradation in a method such as this as amplitude and phase changes cannot be mapped across the baseline. GRAPE uses only a single dimensional parameter space despite genetic algorithms being highly functional at the exploration of high dimensional space \citep{b21,b43}. We are currently implementing an improved algorithm (Bunch of GRAPES) which can stack different combinations of the data points into multiple period spaces simultaneously and use the genetic methods to optimise to a single answer across all the combinations. Alternatively, depending on the combinations of data points stacked by the algorithm, this single answer could instead be a set of results expressed as a function of amplitude and phase allowing for quasi-periodic signals to be expressed in the output. This multi-dimensional approach can also be used to implement a multi-band light curve variant of GRAPE which can determine a set of candidate periods from their simultaneous performance at successfully fitting data in multiple bands. This is accomplished through a modification to the fitness function where the initial single-dimension statistic is weighted against the other dimensions as the chosen candidates should provide satisfactory fits in every band. The Vuong Closeness test can also be modified to evaluate multi-dimensional hyperplanes constructed from multiple sinusoidal models in each dimensional band. Expanding genetic algorithms into multiple period spaces has very interesting applications in upcoming next generation astronomical surveys.

\section*{Acknowledgements}

This research is funded by the Newton Research Collaboration Programme, reference number NRCP1617/5/67.

The Liverpool Telescope is operated on the island of La Palma by Liverpool John Moores University in the Spanish Observatorio del Roque de los Muchachos of the Instituto de Astrofisica de Canarias with financial support from the UK Science and Technology Facilities Council.

We acknowledge with thanks the variable star observations from the AAVSO International Database contributed by observers worldwide and used in this research.









\bsp	
\label{lastpage}
\end{document}